\documentclass[twocolumn,prl,twocolumn,superscriptaddress]{revtex4}
\usepackage[latin9]{inputenc}
\usepackage{color}
\usepackage{amsmath}
\usepackage{amssymb}
\usepackage{graphicx}
\usepackage{tikz}

\usepackage{mathrsfs} 
\usepackage{float}
\usepackage{bbm}

\usepackage{epstopdf}
\usepackage{epsfig} 
\usepackage{verbatim}
\usepackage{array}
\usepackage{setspace}

   \allowdisplaybreaks[1]
   
\usepackage[unicode=true,
 bookmarks=true,bookmarksnumbered=false,bookmarksopen=false,
 breaklinks=false,pdfborder={0 0 1},backref=false,colorlinks=true]
 {hyperref}
\hypersetup{
 linkcolor=magenta, urlcolor=blue, citecolor=blue, pdfstartview={FitH}, hyperfootnotes=false, unicode=true}

\makeatletter
\@ifundefined{textcolor}{}
{%
 \definecolor{BLACK}{gray}{0}
 \definecolor{WHITE}{gray}{1}
 \definecolor{RED}{rgb}{1,0,0}
 \definecolor{GREEN}{rgb}{0,1,0}
 \definecolor{BLUE}{rgb}{0,0,1}
 \definecolor{CYAN}{cmyk}{1,0,0,0}
 \definecolor{MAGENTA}{cmyk}{0,1,0,0}
 \definecolor{YELLOW}{cmyk}{0,0,1,0}
}

\usepackage{amsfonts}\usepackage{tabularx}\usepackage{dcolumn}\usepackage{bm}\usepackage{graphicx}\usepackage{epstopdf}

\hypersetup{urlcolor=blue}

\makeatother

\newcolumntype{C}[1]{>{\centering\arraybackslash$}p{#1}<{$}}

\begin{document}

\title{Leakage and sweet spots in triple-quantum-dot spin qubits: A molecular-orbital study}

\author{Chengxian Zhang}
\affiliation{Department of Physics, City University of Hong Kong, Tat Chee Avenue, Kowloon, Hong Kong SAR, China, and\\
City University of Hong Kong Shenzhen Research Institute, Shenzhen, Guangdong 518057, China}
\author{Xu-Chen Yang}
\affiliation{Department of Physics, City University of Hong Kong, Tat Chee Avenue, Kowloon, Hong Kong SAR, China, and\\
City University of Hong Kong Shenzhen Research Institute, Shenzhen, Guangdong 518057, China}
\author{Xin Wang}
\email{x.wang@cityu.edu.hk}
\affiliation{Department of Physics, City University of Hong Kong, Tat Chee Avenue, Kowloon, Hong Kong SAR, China, and\\
City University of Hong Kong Shenzhen Research Institute, Shenzhen, Guangdong 518057, China}
\date{\today}

\begin{abstract}
A triple-quantum-dot system can be operated as either an exchange-only qubit or a resonant-exchange qubit. While it is generally believed that the decisive advantage of the resonant-exchange qubit is the suppression of charge noise because it is operated at a sweet spot, we show that the leakage is also an important factor. Through molecular-orbital-theoretic calculations, we show that when the system is operated in the exchange-only scheme, the leakage to states with double electron occupancy in quantum dots is severe when rotations around the axis 120$^\circ$ from $\hat{z}$ is performed. While this leakage can be reduced by either shrinking the dots or separating them further, the exchange interactions are also suppressed at the same time, making the gate operations unfavorably slow. When the system is operated as a resonant-exchange qubit, the leakage is 3-5 orders of magnitude smaller. We have also calculated the optimal detuning point which minimizes the leakage for the resonant-exchange qubit, and have found that although it does not coincide with the double-sweet-spot for the charge noise, they are rather close. Our results suggest that the resonant-exchange qubit has another advantage that leakage can be greatly suppressed compared to the exchange-only qubit, and operating at the double-sweet-spot point should be optimal both for reducing charge noise and suppressing leakage.

\end{abstract}

\maketitle

\section{introduction}

A quantum computer is expected to offer exponentially expedited solutions to several important classes of problems compared to the classical ones \cite{nielsen2010quantum}. Semiconductor quantum-dot spin qubits are among the most promising candidates for its realization, partially due to their demonstrated long coherence times and high control fidelities \cite{Petta.05, Bluhm.10b, Barthel.10, Maune.12, Pla.12, Pla.13, Muhonen.14, Kim.14,Kawakami.16}, and partially due to their potential to scale up \cite{Taylor.05}. While the spin states of a single electron naturally constitute a qubit \cite{Loss.98}, difficulties in manipulating them with a time-varying magnetic field \cite{Pioro.05,Koppens.06,Muhonen.14} have led researchers to study spin qubits based on the collective states of two or more electrons. The singlet-triplet qubit \cite{Levy.02} makes use of the two-electron singlet and triplet states, and its manipulation requires control of the Heisenberg exchange interaction between the two spins as well as a static magnetic field gradient. It was then realized that should three spins be used to encode a qubit, the need of a magnetic field gradient can be eliminated, and the control over the exchange interactions suffices for universal operation \cite{DiVincenzo.00,Russ.17}. The triple-quantum-dot three-spin qubit was initially realized as an ``exchange-only'' (EO) qubit \cite{Laird.10}, for which rotations around two axes of the Bloch sphere are achieved by detuning the qubit either positively or negatively toward the spin-to-charge-conversion regimes. Maneuvers of initialization, operation and readout have been experimentally demonstrated \cite{Laird.10, Gaudreau.12,Medford.13a}.

An exchange-only qubit is typically idealized as two $S=1/2$, $S_z=1/2$ (choosing $S_z=-1/2$ is equivalent) states in the three-spin decoherence-free subspace \cite{Fong.11}. Exchange interactions between neighboring spins suffice for arbitrary rotation around the Bloch sphere, with the two rotation axes 120$^\circ$ apart. Another state with the same $S_z$ but $S=3/2$ is energetically close, and the hyperfine noise---the fluctuations of the magnetic field---causes leakage to it \cite{Ladd.12}. Beside the leakage, dephasing can also happen due to either the hyperfine noises \cite{Hung.14,Delbecq.16,Peterfalvi.17} or the charge noises \cite{Eng.15,Thorgrimsson.17}, the latter usually arising from detuning fluctuations caused by unintentionally deposited impurities during the fabrication of the sample. Different from the hyperfine noise, the charge noise is typically dependent on the first-order derivative of the exchange field with respect to the detuning. In order to reduce the charge noise, it is then desirable to operate the qubit at the point where the exchange interaction is first-order insensitive to the detuning, called the ``sweet spot''. This and other considerations have led to the experimental demonstration of the ``resonant-exchange'' (RX) qubit \cite{Medford.13b,Taylor.13,Doherty.13}, for which a triple-spin qubit is operated by radio-frequency (rf) gate voltage pulses around the sweet spot of the exchange interaction. It was later on noted that since there are essentially two independent detuning values, there exists a double sweet spot for the maximal suppression of charge noises \cite{Shim.16,Russ.16}.

While theoretical studies on either the exchange-only or resonant-exchange qubits are abundant, a detailed microscopic theoretical calculation \cite{Burkard.99,Hu.00} of the triple-dot qubits is lacking in the literature. For double quantum dots, the  molecular-orbital-theoretic calculations that are based mostly on configuration interaction method with the Hund-Mulliken approximation, have elucidated many important physics of the system, including for example how the exchange interaction depends on the detailed dimensions of the device and external fields \cite{Burkard.99, Hu.00, He.05, Li.10, Mehl.14}, as well as the response of the device to fluctuations \cite{Nielsen.10, Raith.11, Barnes.11, Bakker.15}.  The vast success originates from the fact that molecular-orbital calculations grant us access to electron wave functions, albeit being approximate, which consequently allows direct calculation of many important physical quantities.  As we shall see in this paper, molecular orbital calculations of a triple-quantum-dot system provides insights to the problem, complementary to theories based on model Hamiltonians. In particular, we found that when the triple-quantum-dot system is operated as an EO qubit, severe leakage will occur when the rotation around the axis 120$^\circ$ from $\hat{z}$ is exercised. While the leakage can be reduced by shrinking the dots or making them further apart from each other, the exchange interaction will also substantially decrease, making the gate operations unacceptably slow. On the other hand, when the triple-quantum-dot system is operated as an RX qubit, the variation of the detuning is confined within the small neighborhood of the sweet spot, which implies a small leakage. Our calculations indicate that the leakage in this case remains substantially smaller than the EO qubit case even when the amplitude of the rf pulse is large. We have also studied how the leakage in the RX qubit depends on the values of detuning, and found that although the point for which the leakage is minimal does not exactly overlap with the double-sweet-spot point, they are rather close.
These results suggest that while the RX qubit is believed to be superior than the EO one due to its suppression of charge noises at the sweet spot, it has another advantage that the leakage can be substantially reduced as compared to the EO qubit.

The remainder of this paper is organized as follows. In Sec.~\ref{sec:model} we explain the model and methods used in our calculations. In Sec.~\ref{sec:res} we present our results, including calculations when the triple-quantum-dot system is operated as an EO qubit (Sec.~\ref{sec:eo}) and as an RX qubit (Sec.~\ref{sec:rx}). In Sec.~\ref{sec:conclusion} we conclude.

\begin{figure}
  \includegraphics[width=0.9\columnwidth]{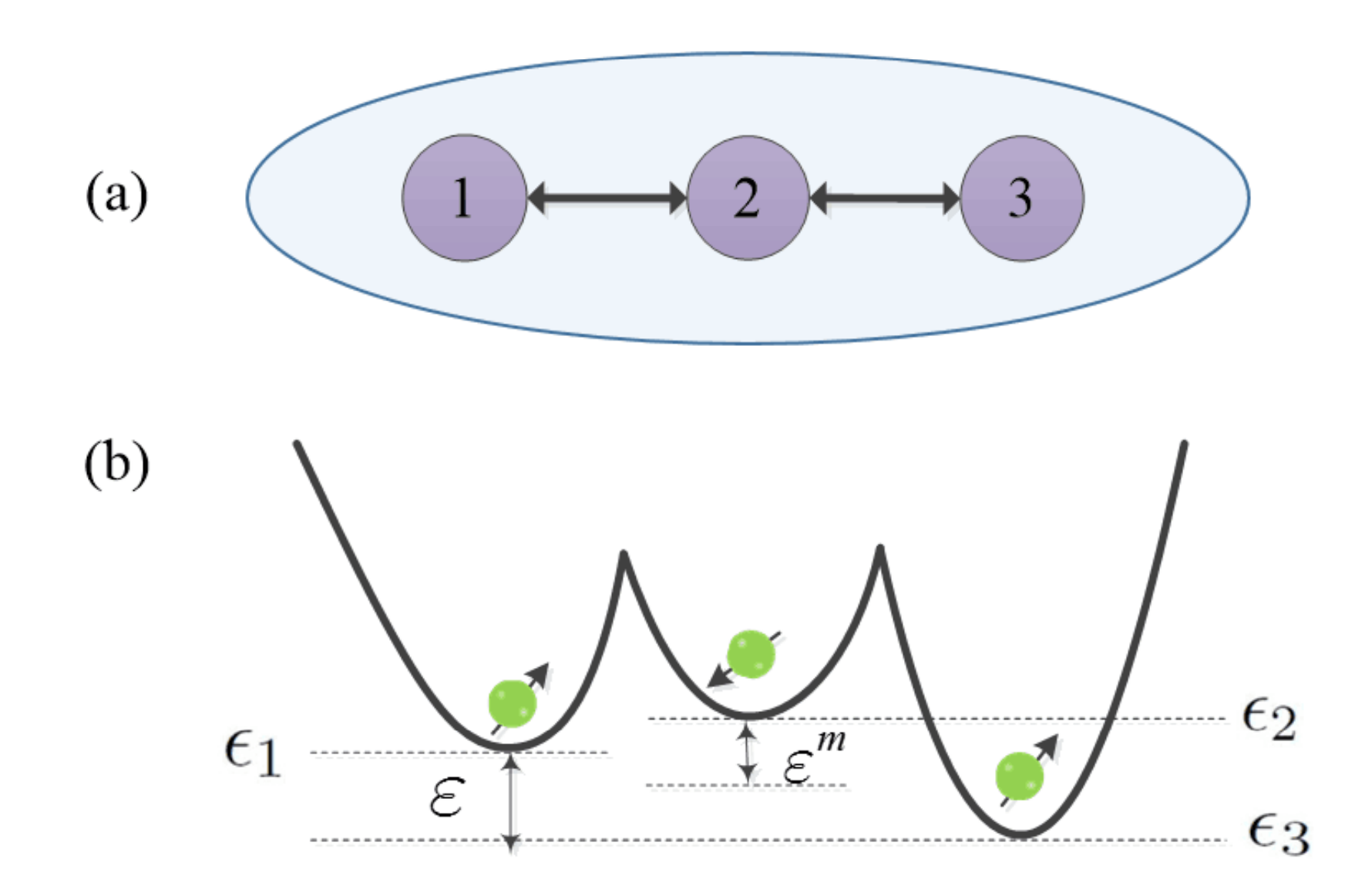}
\caption{(a) Schematic diagram of a lateral triple-quantum-dot system. (b) Schematic of the triple-well confinement potential with the energy of each dot $\epsilon_1$, $\epsilon_2$ and $\epsilon_3$ indicated. The two detuning values $\varepsilon$ and $\varepsilon^m$ are also shown [cf. Eqs.~\eqref{eq:epsl} and \eqref{eq:epslm}].}
\label{fig:intro}
\end{figure}

\section{Model and Methods}
\label{sec:model}

We consider a lateral triple-quantum-dot system with each dot labeled by 1, 2 and 3 from the left to the right [cf. Fig.~\ref{fig:intro}(a)]. When each dot is occupied by a single electron, one may write the effective spin Hamiltonian as
\begin{equation}
H^{\rm spin}=J_{12}\bm{S}_1\cdot\bm{S}_2+J_{23}\bm{S}_2\cdot\bm{S}_3+\sum_{i=1}^3B_iS_i^z,
\label{eq:spinham}
\end{equation}
where $J_{12}$ and $J_{23}$ are Heisenberg exchange interactions between the neighboring spins, and $B_i$ are the magnetic fields at different dots. 
In our model we have neglected the spin-orbit coupling \cite{Stepanenko.03,Rolon.17}. In GaAs, the spin-orbit coupling is very weak \cite{Burkard.99, Stepanenko.12, Cerfontaine.14}, much smaller compared to the level spacing of quantum dot qubits, so it is safe to neglect \cite{Burkard.99,Rancic.14,Cerfontaine.14}. Moreover, the effects of spin-orbit interaction can be minimized by, e.g. reorienting the magnetic fields \cite{Fujita.16,Pal.17,Hofmann.17}, or by applying carefully tailored gates \cite{Bonesteel.01,Burkard.02,Li.17}.

The Hamiltonian \eqref{eq:spinham} has eight eigenstates, out of which the two states with $S=1/2$ and $S_z=1/2$ are chosen as our qubit states $|0\rangle$ and $|1\rangle$ (choosing $S_z=-1/2$ is equivalent but we will stick to the $S_z=1/2$ states throughout this work), and another state with $S=3/2$ and $S_z=1/2$ is the leakage state $|Q\rangle$. These three states can be written as:
\begin{subequations}
\begin{align}
|0_{\rm EO}\rangle&=({|\!\uparrow,\downarrow,\uparrow\rangle}-{|\!\downarrow,\uparrow,\uparrow\rangle})/\sqrt{2},\\
|1_{\rm EO}\rangle&=({|\!\uparrow,\downarrow,\uparrow\rangle}+{|\!\downarrow,\uparrow,\uparrow\rangle})/\sqrt{6}-\sqrt{6}{|\!\uparrow,\uparrow,\downarrow\rangle}/3,\\
|Q\rangle&=({|\!\uparrow,\downarrow,\uparrow\rangle}+{|\!\downarrow,\uparrow,\uparrow\rangle}+{|\!\uparrow,\uparrow,\downarrow\rangle})/\sqrt{3}.
\end{align}
\end{subequations}

When the magnetic field is homogeneous, i.e. $B_1=B_2=B_3$, the Hamiltonian \eqref{eq:spinham} commutes with $\bm{S}^2$ and no leakage occurs. Inhomogeneity of the magnetic field, induced for example by the hyperfine interactions, causes leakage to the $|Q\rangle$ state and is a main source of error \cite{Ladd.12}. This leakage has been studied in the literature and can be suppressed either by enlarging the qubit level spacing \cite{Taylor.13,Hung.14} or by application of dynamically corrected gates \cite{Hickman.13,Zhang.16}. Nevertheless, we would like to focus on the leakage to other states with double electron occupancy, which have received much less attention. We therefore focus on the case when the magnetic field is homogeneous and neglect the hyperfine-coupling-induced inhomogeneity. Here we note that the spin Hamiltonian  \eqref{eq:spinham} is a simplification corresponding to the case in which all dots are occupied by a single electron. Later in this section we will introduce models allowing for double-occupancy and facilitating our discussion of leakage to those states involving doubly occupied dots.

Under the bases of $\{|0_{\rm EO}\rangle,|1_{\rm EO}\rangle,|Q\rangle\}$ (``computational bases''), the spin Hamiltonian \eqref{eq:spinham} can be written in a $3\times3$ matrix \cite{Ladd.12}:
\begin{align}
\begin{split}
H^{\rm comp}_{\rm EO}&=J_{12}E_{12}+J_{23}E_{23}\\
&+\left(\frac{\lambda_1}{2 \sqrt{3}}+\frac{\lambda_4}{\sqrt{6}}\right)\Delta_A+\left(\frac{\lambda_3}{3}+\frac{\sqrt{2}}{3}\lambda_6\right)\Delta_B,
\end{split}
\label{eq:compham}
\end{align}
where
\begin{align}
E_{12}=-\frac{\lambda_3}{2}-\frac{\lambda_8}{2\sqrt{3}}, \quad E_{23}=-\frac{\sqrt{3}}{4}\lambda_1+\frac{\lambda_3}{4}-\frac{\lambda_8}{2\sqrt{3}},
\end{align}
$\lambda_i$ are Gell-Mann matrices \cite{Georgi.99},  
$\Delta_A=B_1-B_2$ and $\Delta_B=B_3-(B_1+B_2)/2$ are fluctuations in the hyperfine field \cite{Ladd.12,Hickman.13} which cause, in this case, leakage to $|Q_{\rm EO}\rangle$. In absence of leakage, $E_{12}$ and $E_{23}$ implement rotations on the Bloch sphere around two axes which are 120$^\circ$ apart, i.e. $\hat{z}$ and $\frac{\sqrt{3}}{2}\hat{x}-\frac{1}{2}\hat{z}$, respectively. Therefore the control over the two axes suffices for arbitrary single-qubit rotations. 

The spin Hamiltonian \eqref{eq:spinham} is a simplification of the problem, which assumes that the three quantum dots are occupied by one electron each. Moreover, this theory alone cannot provide the value of exchange interactions from the microscopic details (i.e. detuning, dot sizes and distance between the dots). Therefore we also need a microscopic Hamiltonian which explicitly allows double occupancy in a given dot. We take the magnetic field to be along the $z$ direction and the quantum dots within the $xy$ plane. The microscopic Hamiltonian can be written as
\begin{equation}
H^{\rm micro}=\sum_{i=1}^3\left[\frac{1}{2m^*}\left(\bm{p}_i-e\bm{A}_i\right)^2+V(\bm{r}_i)\right],
\label{eq:H}
\end{equation}
where $m^*$ is the effective mass of the electron (taken to be 0.067 electron mass for GaAs), $V(\bm{r})$ is the confinement potential of the triple quantum dots. 

In this work $V(\bm{r})$ is modeled as 
\begin{equation}
\begin{aligned}
V(\bm{r})=\text{Min}\left[v_1(\bm{r}),v_2(\bm{r}),v_3(\bm{r})\right],
\end{aligned}
\label{eq:V}
\end{equation}
where
\begin{equation}
\begin{aligned}
v_i(\bm{r})\equiv\frac{m^*\omega_0^2}{2}\left|\bm{r}-\bm{R}_i\right|^2+\epsilon_i
\end{aligned}
\label{eq:v}
\end{equation}
is the confinement potential for the $i$th quantum dot, which  centers at $\bm{R}_i$ $(i=1,2,3)$: 
\begin{align}
\bm{R}_1=(-2a,0),\quad\bm{R}_2=(0,0),\quad\bm{R}_3=(2a,0).
\label{eq:R}
\end{align}
A schematic diagram of the potential is shown in Fig.~\ref{fig:intro}(b).

Here, the key parameters describing the microscopic details of the systems are $\omega_0$, the confinement energy \cite{Burkard.99, Hu.00} characterizing the size of the dot, and $\epsilon_i$, the energies of electrons in each dot. Differences in $\epsilon_i$ are termed as the ``detuning'' which can be used to vary the amplitudes of $J_{12}$ and $J_{23}$. There are two independent detunings, $\varepsilon$ and $\varepsilon^m$ \cite{Shim.16,Russ.16}, defined as
\begin{subequations}
\begin{align}
\varepsilon&=\epsilon_1-\epsilon_3,\label{eq:epsl}\\
\varepsilon^m&=\epsilon_2-(\epsilon_1+\epsilon_3)/2.\label{eq:epslm}
\end{align}
\end{subequations}
Typically the EO qubit is manipulated by varying $\varepsilon$ with $\varepsilon^m$ fixed. However, since $J_{12}$ and  $J_{23}$ are dependent on both $\varepsilon$ and $\varepsilon^m$, in order to suppress the charge noise one may define a ``double sweet spot'' with $\partial J_{12}/\partial\varepsilon=
\partial J_{12}/\partial\varepsilon^m=
\partial J_{23}/\partial\varepsilon=
\partial J_{23}/\partial\varepsilon^m=0$, and operate the qubit close to the sweet spot \cite{Medford.13b,Shim.16}. This has inspired the invention of the RX qubit, which is operated by an rf pulse in a small neighborhood of the sweet spot.  
In the initial demonstration of the RX qubit, it is operated at the sweet spot of $\varepsilon$ only with $\varepsilon^m$ assumed to be fixed at some value \cite{Medford.13b}. Ref.~\cite{Shim.16} pointed out that one should consider the double sweet spot (namely, the ``three-dimensional sweet spot'') as explained above. Very recently, there has been experimental demonstration of operations of an RX qubit at this double sweet spot 
\cite{Malinowski.17}. While no significant improvement of the gate fidelity is found, more investigations are needed along this line.

 In this paper, we are interested in properties of triple-quantum-dot qubits (including both EO and RX qubits) which are calculated from the microsopic theory.  We will show that beside the fact that the RX qubit has smaller charge noise than the EO one, it also has another advantage: the leakage to other states, which are significant in EO, is also substantially suppressed if the triple-quantum-dot system is treated as an RX qubit. To do this we need access to the detailed energy level structure of the system which can only be made available from a microscopic calculation.

To solve this multi-electron problem, we use the configuration interaction method building electron wave functions from the Fock-Darwin states, which are essentially the harmonic oscillator states. We
adopt the Hund-Mulliken approximation \cite{Burkard.99,Hu.00} that retains only the ground states
\begin{equation}
\begin{aligned}
\phi_i(\bm{r})=\frac{1}{a_B^{}\sqrt{\pi}}\text{exp}\left[{-\frac{1}{2a_B^2}\left|\bm{r}-\bm{R}_i\right|^2}\right],\quad i=1,2,3.
\end{aligned}
\label{eq:ground}
\end{equation}
Here, $a_B^{}\equiv\sqrt{\hbar/m\omega_0}$ is Fock-Darwin radius. 
A set of orthogonal single-electron states
can then be built   by the transformation
\begin{equation}
\begin{aligned}
\left\{\psi_1,\text{ }\psi_2,\text{ }\psi_3\right\}^\text{T}=\mathcal{O}^{-1/2}\left\{\phi_1,\text{ }\phi_2,\text{ }\phi_3\right\}^\text{T},
\end{aligned}
\label{eq:trans}
\end{equation}
where $\mathcal{O}$ is the overlap matrix defined as $\mathcal{O}_{l,l^\prime}\equiv\langle\phi_l|\phi_{l^\prime}\rangle$ and found following \cite{annavarapu2013singular,Yang.17}.

The Hamiltonian of the three-electron system can also be expressed, using the second-quantized notations, in a Hubbard-like form as \cite{Yang.11,Yang.17}
\begin{equation}
H^{\rm Hubbard}=H_e+H_t+H_U+H_{J^e}+H_{J^p}+H_{J^t},
\label{eq:fqdHam}
\end{equation}
where
\begin{subequations}
\begin{align}
H_e=&\sum^3_{k=1}\sum_\sigma\epsilon_kc^\dagger_{k\sigma}c^{}_{k\sigma},\label{eq:A6a}\\
H_t=&\sum^{2}_{k=1}\sum_\sigma t_{k,k+1}c^\dagger_{k\sigma}c^{}_{k+1,\sigma}+\text{H.c.},\label{eq:A6b}\\
H_U=&\sum^2_{k=1}U_{k,k+1}\left(n_{k\uparrow}+n_{k\downarrow}\right)\left(n_{k+1,\uparrow}+n_{k+1,\downarrow}\right)\notag\\
&+\sum^3_{k=1}U_kn_{k\uparrow}n_{k\downarrow},\label{eq:A6c}\\
H_{J^e}=&-\sum^2_{k=1}\sum_{\sigma_1,\sigma_2}J^e_{k,k+1}c^\dagger_{k\sigma_1}c^\dagger_{k+1,\sigma_2}c^{}_{k+1,\sigma_1}c^{}_{k\sigma_2},\label{eq:A6d}\\
H_{J^p}=&-\sum^2_{k=1}J^p_{k,k+1}c^\dagger_{k+1,\uparrow}c^\dagger_{k+1,\downarrow}c_{k\uparrow}c_{k\downarrow}+ \text{H.c.},\label{eq:A6e}\\
H_{J^t}=&-\sum^2_{k=1}\sum^{k+1}_{i=k}\sum_\sigma J^{t,i}_{k,k+1}n_{i\sigma}c^\dagger_{k\bar{\sigma}}c_{k+1,\bar{\sigma}}+\text{H.c.}.
\label{eq:A6f}
\end{align}
\end{subequations}
Here, the $c^\dagger_{i\sigma}$ operator creates  an electron with spin $\sigma$ at the $i^{\rm th}$ dot.

In our problem,  three electrons (two spin-up and one spin-down) occupy the lateral triple-quantum-dot system, and each dot allows a maximum of two electrons. There are a total of 9 possibilities so our complete bases contain the following 9 states:
\begin{subequations}
		\begin{align}
	\left | \uparrow,\uparrow,\downarrow \right \rangle&=c^\dagger_{1\uparrow}c^\dagger_{2\uparrow}c^\dagger_{3\downarrow}|\text{vac}\rangle,\label{eq:9states1}\\
	\left | \uparrow,\downarrow,\uparrow \right \rangle&=c^\dagger_{1\uparrow}c^\dagger_{2\downarrow}c^\dagger_{3\uparrow}|\text{vac}\rangle,\\
	\left | \downarrow,\uparrow,\uparrow \right \rangle&=c^\dagger_{1\downarrow}c^\dagger_{2\uparrow}c^\dagger_{3\uparrow}|\text{vac}\rangle,\\	
	\left | \uparrow,\uparrow\downarrow,0 \right \rangle&= c^\dagger_{1\uparrow}c^\dagger_{2\uparrow}c^\dagger_{2\downarrow}|\text{vac}\rangle,\\
	\left | \uparrow\downarrow,\uparrow,0 \right \rangle&= c^\dagger_{1\uparrow}c^\dagger_{1\downarrow}c^\dagger_{2\uparrow}|\text{vac}\rangle,\\
	\left |0, \uparrow,\uparrow\downarrow \right \rangle&= c^\dagger_{2\uparrow}c^\dagger_{3\uparrow}c^\dagger_{3\downarrow}|\text{vac}\rangle,\\
	\left |0, \uparrow\downarrow,\uparrow \right \rangle&= c^\dagger_{2\uparrow}c^\dagger_{2\downarrow}c^\dagger_{3\uparrow}|\text{vac}\rangle,\\
	\left | \uparrow,0,\uparrow\downarrow \right \rangle&= c^\dagger_{1\uparrow}c^\dagger_{3\uparrow}c^\dagger_{3\downarrow}|\text{vac}\rangle,\\
	\left | \uparrow\downarrow,0,\uparrow \right \rangle&= c^\dagger_{1\uparrow}c^\dagger_{1\downarrow}c^\dagger_{3\uparrow}|\text{vac}\rangle\label{eq:9states9},
		\end{align}		
\end{subequations}
where $|\text{vac}\rangle$ refers to a vacuum state. To facilitate discussions in the remainder of this paper, we also introduce the notation $(n_1,n_2,n_3)$ to be used interchangeably with the l.h.s.~of Eqs.~\eqref{eq:9states1}-\eqref{eq:9states9}, where $n_i$ denotes the electron occupancy of the $i$th dot, $i=1,2,3$. For example, $(1,1,1)$ refers to any linear superposition of $\left | \uparrow,\uparrow,\downarrow \right \rangle$, $\left | \uparrow,\downarrow,\uparrow \right \rangle$ and $\left | \downarrow,\uparrow,\uparrow \right \rangle$ states, $(1,0,2)$ refers to $\left | \uparrow,0,\uparrow\downarrow \right \rangle$ and $(2,0,1)$ the $\left | \uparrow\downarrow,0,\uparrow \right \rangle$ state. It is obvious that $(1,1,1)$ indicates the computational subspace of the qubit, and we shall see in Sec.~\ref{sec:eo} that the leakage to $(2,0,1)$ and $(1,0,2)$ can seriously hinder the coherent operation of the system as an  EO qubit, while the leakage is substantially reduced when the system is operated as an RX qubit.

Under the bases Eqs.~\eqref{eq:9states1}-\eqref{eq:9states9}, the Hubbard Hamiltonian \eqref{eq:fqdHam} is written in a $9\times9$ matrix. $H_e$ (the kinetic energy), $H_U$ (Coulomb repulsions) constitute the diagonal elements of the matrix, while $H_t$ (hopping), $H_{J^e}$ (spin super-exchange), $H_{J^p}$ (pair-hopping) and $H_{J^t}$  (occupation-modulated hopping) terms contribute to the off-diagonal ones.

The configuration interaction calculation is essentially the evaluation of the Hubbard parameters involved in Eq.~\eqref{eq:fqdHam} using inner products of the orthogonalized electron wave functions built on the results of Eq.~\eqref{eq:trans}. The energy spectra and the composition of eigenstates are then found by diagonalization of the matrix. 

\begin{figure}
	\includegraphics[width=0.8\columnwidth]{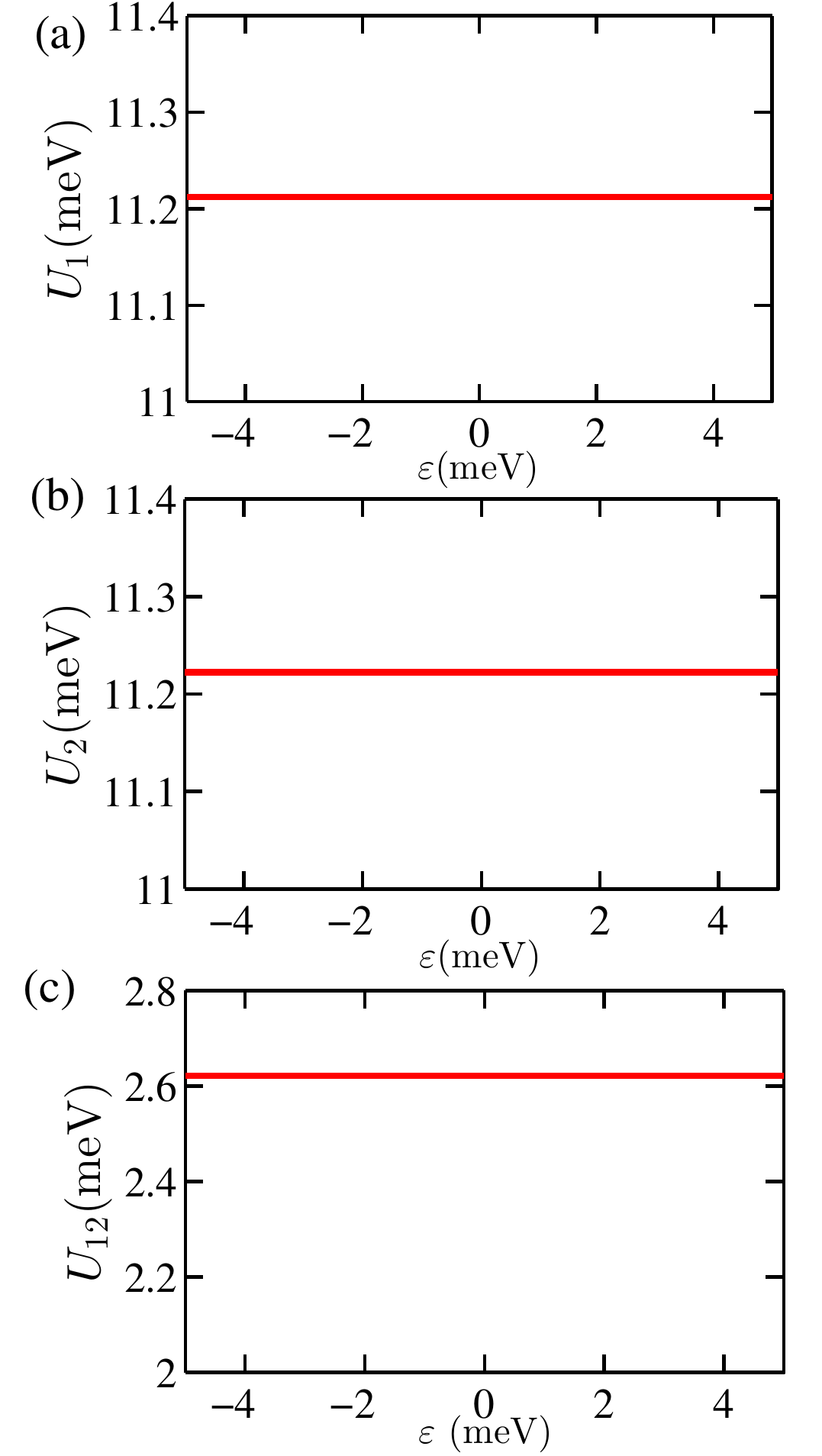}
	\caption{Calculated parameters $U_{1}$, $U_{2}$ and $U_{12}$ in the Hamiltonian, Eq.~\eqref{eq:A6c}, as functions of $\varepsilon$. Parameters: $\hbar\omega_{0}=8 \mathrm{meV}$, $a=22 \mathrm{nm}$ and $\varepsilon^{m}=3.3 \mathrm{meV}$.}
	\label{fig:Uplot}
\end{figure}

Figure~\ref{fig:Uplot} shows the calculated $U_{1}$, $U_{2}$ and $U_{12}$ in the Hamiltonian, Eq.~\eqref{eq:A6c}, as functions of $\varepsilon$. Because of the symmetry of the triple-well potential, $U_1=U_3$ and $U_{12}=U_{23}$. Because the Coulomb interaction essentially depends on the occupancy of each dot, it almost does not change with detuning, as has been shown in Fig.~\ref{fig:Uplot}. The results shown in Fig.~\ref{fig:Uplot} are obtained for $\varepsilon^{m}=3.3 \mathrm{meV}$, but there is almost no change when $\varepsilon^{m}$ becomes negative (unlike the case for the inter-dot hopping to be explained below) so the results for $\varepsilon^{m}<0$ are not shown. From Fig.~\ref{fig:Uplot} we also see that the inter-dot Coulomb interaction $U_{12}$ is much weaker than the intra-dot ones, $U_1$ and $U_2$, as expected from the overlap between the electron wave functions.

\begin{figure}
	\includegraphics[width=.9\columnwidth]{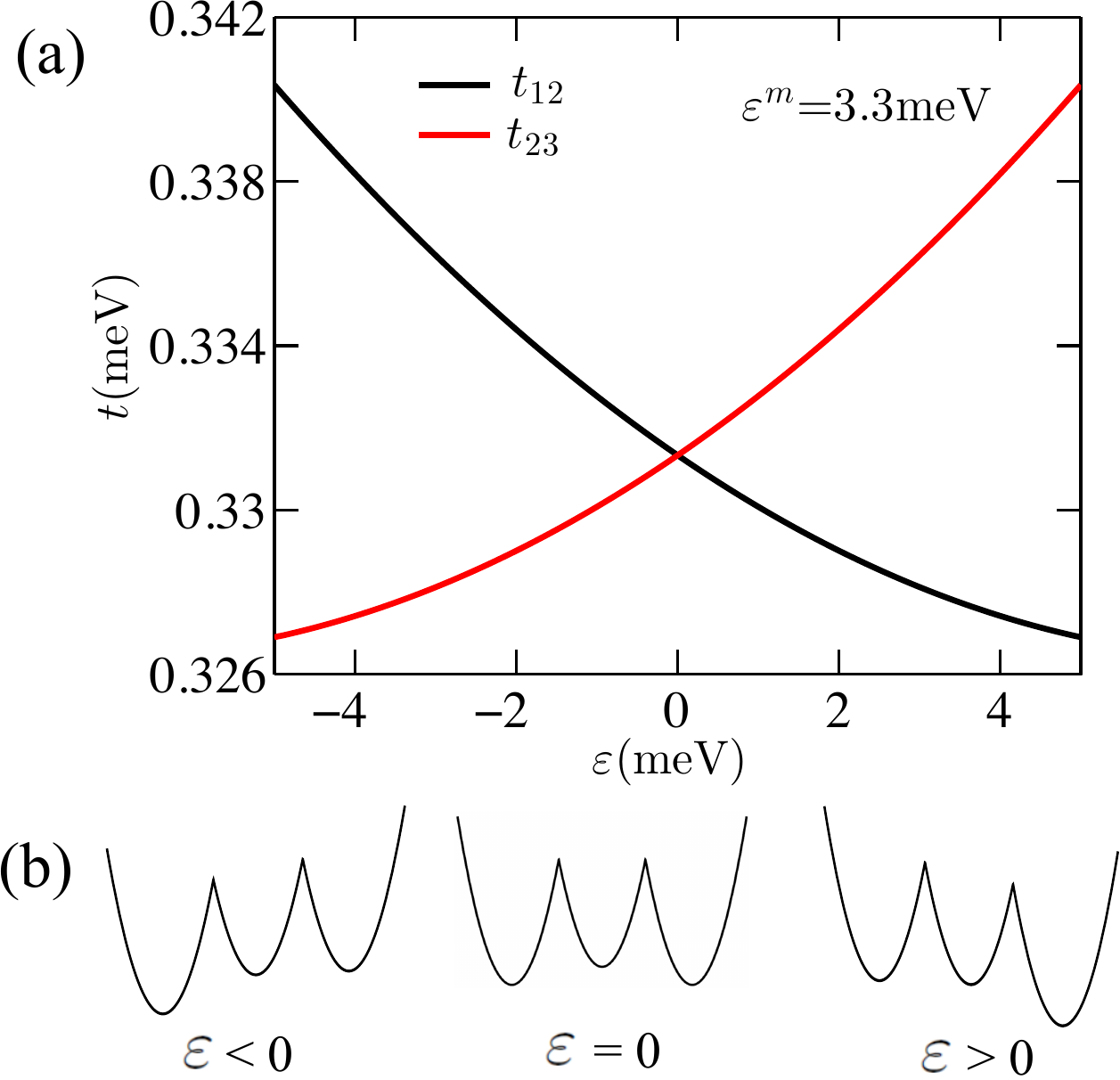}
	\caption{(a) Calculated parameters $t_{12}$ (black line) and $t_{23}$ (red/gray line) in the Hamiltonian, Eq.~\eqref{eq:A6b}, as functions of $\varepsilon$ for $\varepsilon^{m}=3.3 \mathrm{meV}$. (b) Schematic plots of the  triple-well confinement potential with detuning $\varepsilon<0$, $\varepsilon=0$ and $\varepsilon>0$ as indicated. Parameters: $\hbar\omega_{0}=8 \mathrm{meV}$, $a=22 \mathrm{nm}$.}
	\label{fig:eo}
\end{figure}

\begin{figure}
	\includegraphics[width=.9\columnwidth]{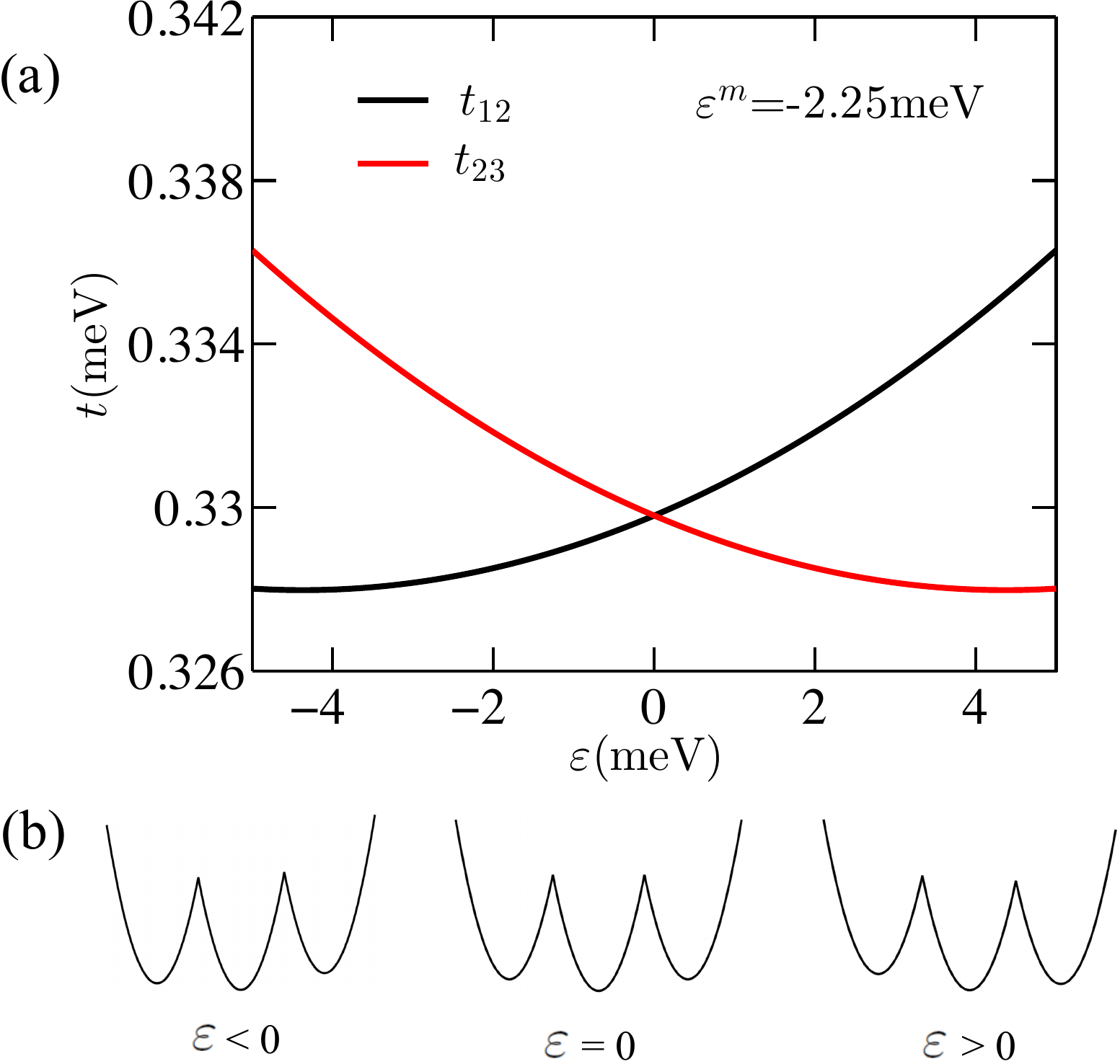}
	\caption{(a) Calculated parameters $t_{12}$ (black line) and $t_{23}$ (red/gray line) in the Hamiltonian,  Eq.~\eqref{eq:A6b}, as functions of $\varepsilon$ for $\varepsilon^{m}=-2.25 \mathrm{meV}$. (b) Schematic plots of the triple-well confinement potential with detuning $\varepsilon<0$, $\varepsilon=0$ and $\varepsilon>0$ as indicated. Parameters: $\hbar\omega_{0}=8 \mathrm{meV}$, $a=22 \mathrm{nm}$.}
	\label{fig:rx}
\end{figure}

In Fig.~\ref{fig:eo} and Fig.~\ref{fig:rx} we show the calculated inter-dot hopping $t_{12}$ and $t_{23}$ as functions of the detuning $\varepsilon$, for two values of $\varepsilon^m$: Fig.~\ref{fig:eo} shows the results for $\varepsilon^{m}=3.3 \mathrm{meV}$ which corresponds to our choice of parameters for the discussion of the EO qubit; Fig.~\ref{fig:rx} shows the  $\varepsilon^{m}=-2.25 \mathrm{meV}$ case which is at the double sweet spot of an RX qubit. For $\varepsilon^{m}>0$, we see from Fig.~\ref{fig:eo}(a) that $t_{12}$ drops as $\varepsilon$ turns from a negative value to a positive one, while $t_{23}$ increases. Fig.~\ref{fig:eo}(b) shows the schematic plots of the confinement potential. When $\varepsilon=0$, the potential well in both dots 1 and 3 are deeper than that of dot 2. As $\varepsilon$ is turned to negative values, dot 1 becomes even deeper, therefore the electron in dot 2 is more likely to move to dot 1, resulting in an increased $t_{12}$. On the other hand, as the energy of dot 1 is decreased, that of dot 3 is raised, so that the energies of dots 2 and 3, $\epsilon_2$ and $\epsilon_3$ become comparable, therefore $t_{23}$ is decreased. The opposite happens when $\varepsilon$ is turned to positive values. 

On the contrary, on the double sweet spot of the RX qubit, $\varepsilon^{m}<0$. From Fig.~\ref{fig:rx} we see that $t_{12}$ decreases as $\varepsilon$ is turned from negative to positive values, while $t_{23}$ increases. This is because in this case, $\epsilon_2<\epsilon_1,\epsilon_3$ when $\varepsilon=0$ [cf. Fig.~\ref{fig:rx}(b)], and even if the system is detuned a little, an electron is not likely to move from the middle dot to its sides. On the contrary, when $\varepsilon<0$, an electron in dot 3 will be more likely to move to dot 2, so $t_{23}$ is larger. The electron in dot 1 is less likely to move to dot 2 since  $\epsilon_1$ and $\epsilon_2$ become comparable, resulting in a smaller $t_{12}$. The opposite is true when $\varepsilon>0$. As we shall see in Sec.~\ref{sec:rx}, these results have interesting consequences on the leakage. We also note here that the exchange interactions ($J^e$, $J^p$, $J^t$) in Eqs.~\eqref{eq:A6d}, \eqref{eq:A6e} and \eqref{eq:A6f} are much smaller so we do not show the results for them here.

\section{Results}
\label{sec:res}

\subsection{Exchange-only qubit}
\label{sec:eo}

\begin{figure}
	\includegraphics[width=0.85\columnwidth]{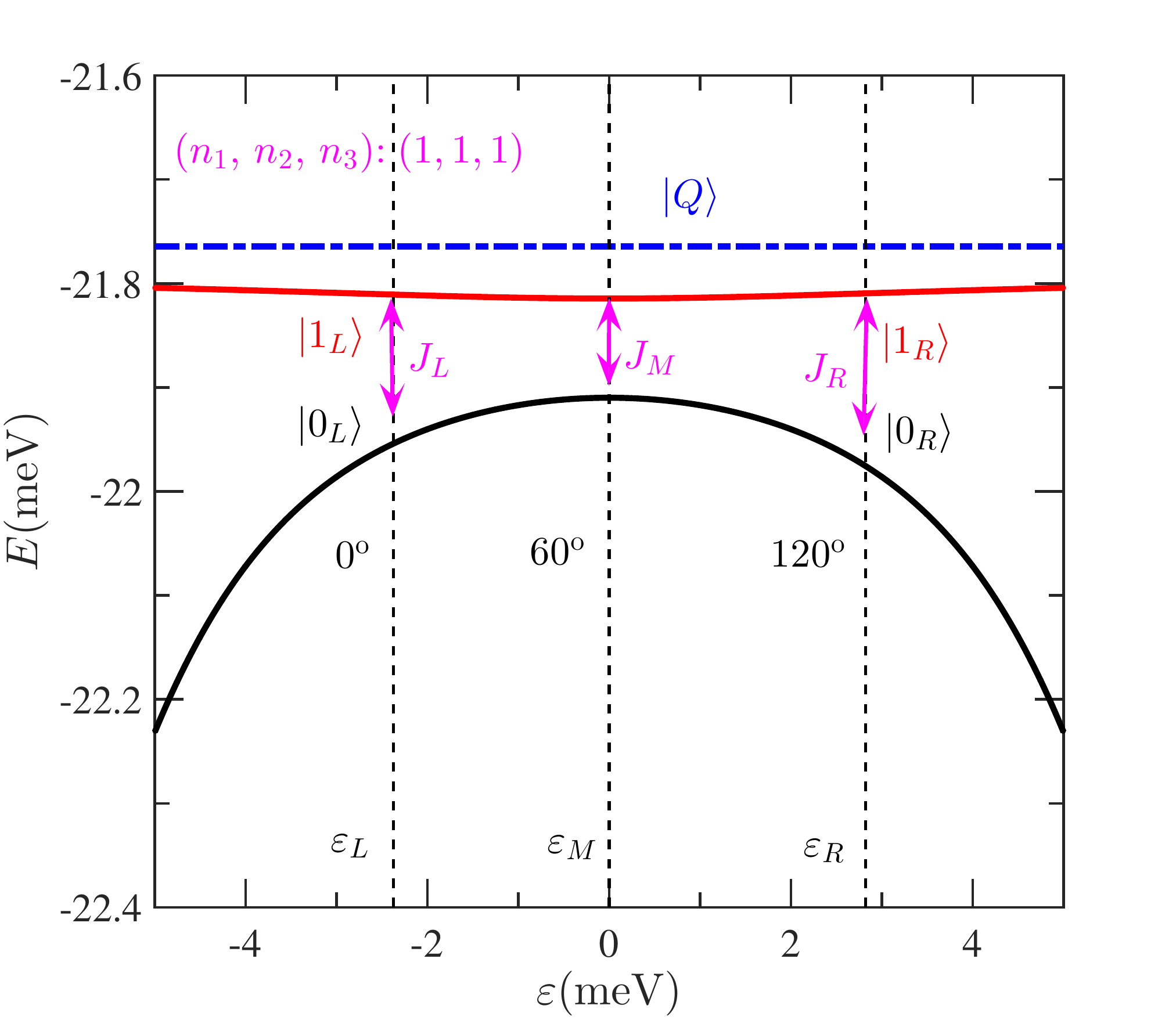}
	\caption{Calculated energy spectra of the triple-quantum dot system.  The ground state is marked as $\left|\mathrm{GS}\right\rangle$, the first excited state as  $\left|\mathrm{1ES}\right\rangle$ and the second excited state  $\left | Q \right \rangle$. The splitting between  $\left|\mathrm{GS}\right\rangle$ and $\left|\mathrm{1ES}\right\rangle$ gives the exchange interactions which, at different $\varepsilon$ values, are denoted as $J_{L}$, $J_{M}$ and $J_{R}$. The angles near the vertical dashed lines give the direction of the rotating axis if a qubit is operated at the corresponding detuning. Parameters: $\hbar\omega_{0}=8 \mathrm{meV}$, $a=22 \mathrm{nm}$ and $\varepsilon^{m}= 3.3 \mathrm{meV}$. (Note that one should not confuse $\varepsilon_M$ with $\varepsilon^m$: $\varepsilon_M=0$ in this figure is the detuning value of $\varepsilon$ relevant to $\epsilon_1$ and $\epsilon_3$ at which the exchange only qubit is rotated at an axis 60$^\circ$ apart from $\hat{z}$ (cf. Eq.~\eqref{eq:epsl}), while $\varepsilon^m$ is another detuning value defined in Eq.~\eqref{eq:epslm}.) }
	\label{fig:5}
\end{figure}

\begin{figure}
\includegraphics[width=0.71\columnwidth]{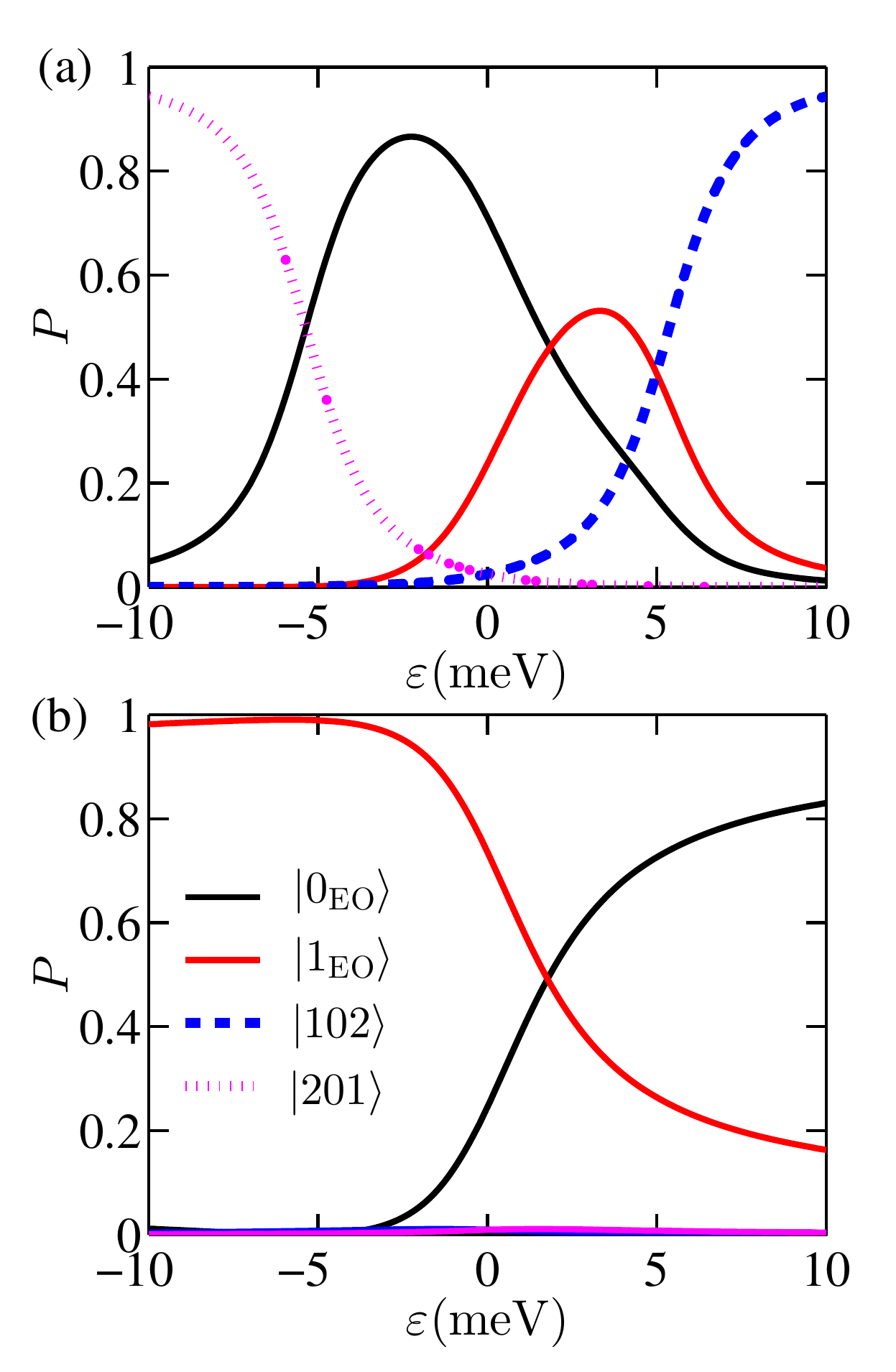}
\caption{The composition of states as functions of the detuning $\varepsilon$. (a): The composition of the ground state $\left|\mathrm{GS}\right\rangle$. (b): The composition of the first excited state  $\left|\mathrm{1ES}\right\rangle$. Parameters: $\hbar\omega_{0}=8 \mathrm{meV}$, $a=22 \mathrm{nm}$ and $\varepsilon^{m}=3.3 \mathrm{meV}$.}
\label{fig:6}
\end{figure}

In this section, we study the energy-level spectra of the triple-quantum-dot system. The energy levels can be tuned by changing the detuning value $\varepsilon$ which subsequently varies the charge configuration. There are a total of 9 energy levels in our problem (cf. Eqs.~\eqref{eq:9states1}-\eqref{eq:9states9}), but we only study the three most relevant ones as depicted in Fig.~\ref{fig:5}. In Fig.~\ref{fig:5}, the black solid line shows the ground state ($\left|\mathrm{GS}\right\rangle$), the red(gray) solid line the first excited state ($\left|1\mathrm{ES}\right\rangle$), and the blue(gray) dash-dotted line the second excited state which is denoted by $\left|Q\right\rangle$ because it is exclusively composed of the $\left|Q\right\rangle$  state. The splitting between $\left|\mathrm{GS}\right\rangle$ and $\left|1\mathrm{ES}\right\rangle$ gives the exchange interaction. The exchange interaction at three points has been marked for the convenience of discussions later:  $J_L$ at $\varepsilon=\varepsilon_L= -2.37$ meV, $J_M$ at $\varepsilon=\varepsilon_M=0$, and $J_R$ at $\varepsilon=\varepsilon_R= 2.82$ meV. The corresponding ground states (first excited states) are termed as $|0_L\rangle$ ($|1_L\rangle$), $|0_M\rangle$ ($|1_M\rangle$), and $|0_R\rangle$ ($|1_R\rangle$). When the triple-quantum-dot system is operated as an exchange-only qubit encoded as $|0_\mathrm{EO}\rangle=|0_L\rangle$ and $|1_\mathrm{EO}\rangle=|1_L\rangle$, the exchange interaction constitutes the rotation around the $\hat{z}$ axis (0$^\circ$). For $\varepsilon=\varepsilon_M$ and $\varepsilon=\varepsilon_R$ the corresponding exchange interaction rotates the Bloch vector around the axes in the $xz$ plane that are 60$^\circ$ and 120$^\circ$ apart from $\hat{z}$, respectively. Throughout this paper we fix the external magnetic field at zero so that there is no leakage to the $|Q\rangle$ state. We then focus on the leakage to other states which involves double electron occupancy.

The detuning value $\varepsilon$ is varied from $-5 \mathrm{meV}$ to $5 \mathrm{meV}$ in Fig.~\ref{fig:5}, in which the ground states are mostly composed by the  (1,1,1) charge configuration. For $\varepsilon\lesssim -5\mathrm{meV}$, (2,0,1) is the dominating configuration of the ground state, while for $\varepsilon\gtrsim5\mathrm{meV}$  (1,0,2) dominates. This can be clearly seen in Fig.~\ref{fig:6} where the compositions of the ground state (panel (a)) and the first excited state (panel (b)) are shown. From Fig.~\ref{fig:6}(a) we can see that the composition (probability) of the $|0_\mathrm{EO}\rangle$ state has a peak at $\varepsilon= -2.37$ meV which we have defined as $\varepsilon_L$. As $\varepsilon$ increases, the character of $|0_\mathrm{EO}\rangle$  decreases and that of $|1_\mathrm{EO}\rangle$ increases, the latter of which peaks at $\varepsilon\approx3$ meV (note that $\varepsilon_R$ is defined not exactly at this point---although very close---but instead, at the point where the exchange interaction performs a rotation around an axis 120$^\circ$ from $\hat{z}$). For $\varepsilon\lesssim -7$ meV ($\varepsilon\gtrsim7$ meV) the character of $|201\rangle$ ($|102\rangle$) exceeds 80\%. From Fig.~\ref{fig:6}(b) we see that the admixture of $|201\rangle$ and $|102\rangle$ are both negligibly small and for $\varepsilon\lesssim1$ meV ($\varepsilon\gtrsim1$ meV) the $|1_\mathrm{EO}\rangle$ ($|0_\mathrm{EO}\rangle$) state dominates. 

When operating this triple-quantum-dot system, we define the ground state $|0_L\rangle$ and the first excited state $|1_L\rangle$  at $\varepsilon=\varepsilon_L= -2.37$ meV as our logical 0 and 1, respectively. Note the two states are not exclusively $|0_\mathrm{EO}\rangle$ or $|1_\mathrm{EO}\rangle$, but are rather linear superposition of all the 9 states involved in Eqs.~\eqref{eq:9states1}-\eqref{eq:9states9}. In operating the exchange-only qubit, the detuning is rapidly pulsed to other values so that the states at $\varepsilon\neq\varepsilon_L$ must be decomposed into the states of $|0_L\rangle$ and $|1_L\rangle$. This has two consequences: first, decomposing to the subspace spanned by $|0_L\rangle$ and $|1_L\rangle$ means that the rotation caused by the exchange interaction at that point should be around an axis apart from $\hat{z}$, and the angle can be calculated from the composition of $|0_L\rangle$ and $|1_L\rangle$ states. Second, since we have a total of 9 bases, and $|0_L\rangle$ and $|1_L\rangle$ are merely two of them, this projection inevitably involves leakage, i.e. the total probability of measuring either  $|0_L\rangle$ or  $|1_L\rangle$ for the $|\mathrm{GS}\rangle$ and $|\mathrm{1ES}\rangle$ states concerned  will be less than 100\%. The difference from 100\% is thus defined as ``leakage'', denoted by $\eta$ because it is not involved in the computation subspace. $\eta=0$ for $\varepsilon=\varepsilon_L$ but increases as $\varepsilon$ is shifted away from $\varepsilon_L$.

\begin{figure}
	\includegraphics[width=0.9\columnwidth]{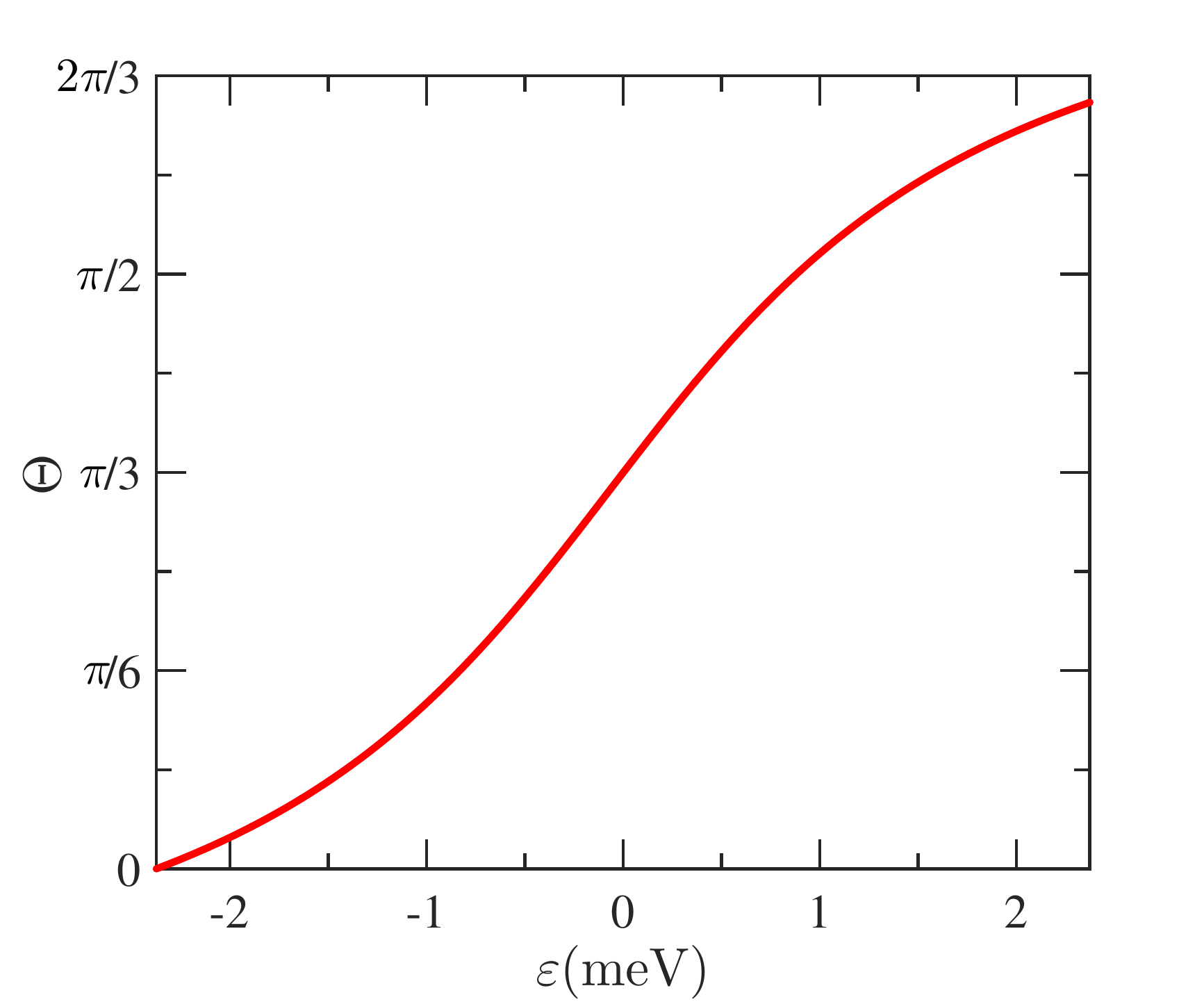}
	\caption{The polar angle (the angle between the rotation axis and $\hat{z}$) v.s. detuning $\varepsilon$, when $|0_L\rangle$ and $|1_L\rangle$ are defined as qubit states. Parameters: $\hbar\omega_0=8 \mathrm{meV}$, $a=22 \mathrm{nm}$ and $\varepsilon^{m}=3.3 \mathrm{meV}$.}
	\label{fig:7}
\end{figure}

Fig.~\ref{fig:7} shows the polar angle $\Theta$, the angle between the rotation axis and $\hat{z}$, at different detuning $\varepsilon$ values. $\Theta=0$ at $\varepsilon=\varepsilon_L= -2.37$ meV and increases as $\varepsilon$ is increased. $\Theta=\pi/3$ (60$^\circ$) at $\varepsilon=\varepsilon_M=0$, and we have found that $\Theta=2\pi/3$ (120$^\circ$) at $\varepsilon=\varepsilon_R=2.82$ meV for the system concerned. This asymmetry that $|\varepsilon_R|\neq|\varepsilon_L|$ is due to leakage out of the space spanned by $|0_L\rangle$ and $|1_L\rangle$.

\begin{figure}
\includegraphics[width=0.85\columnwidth]{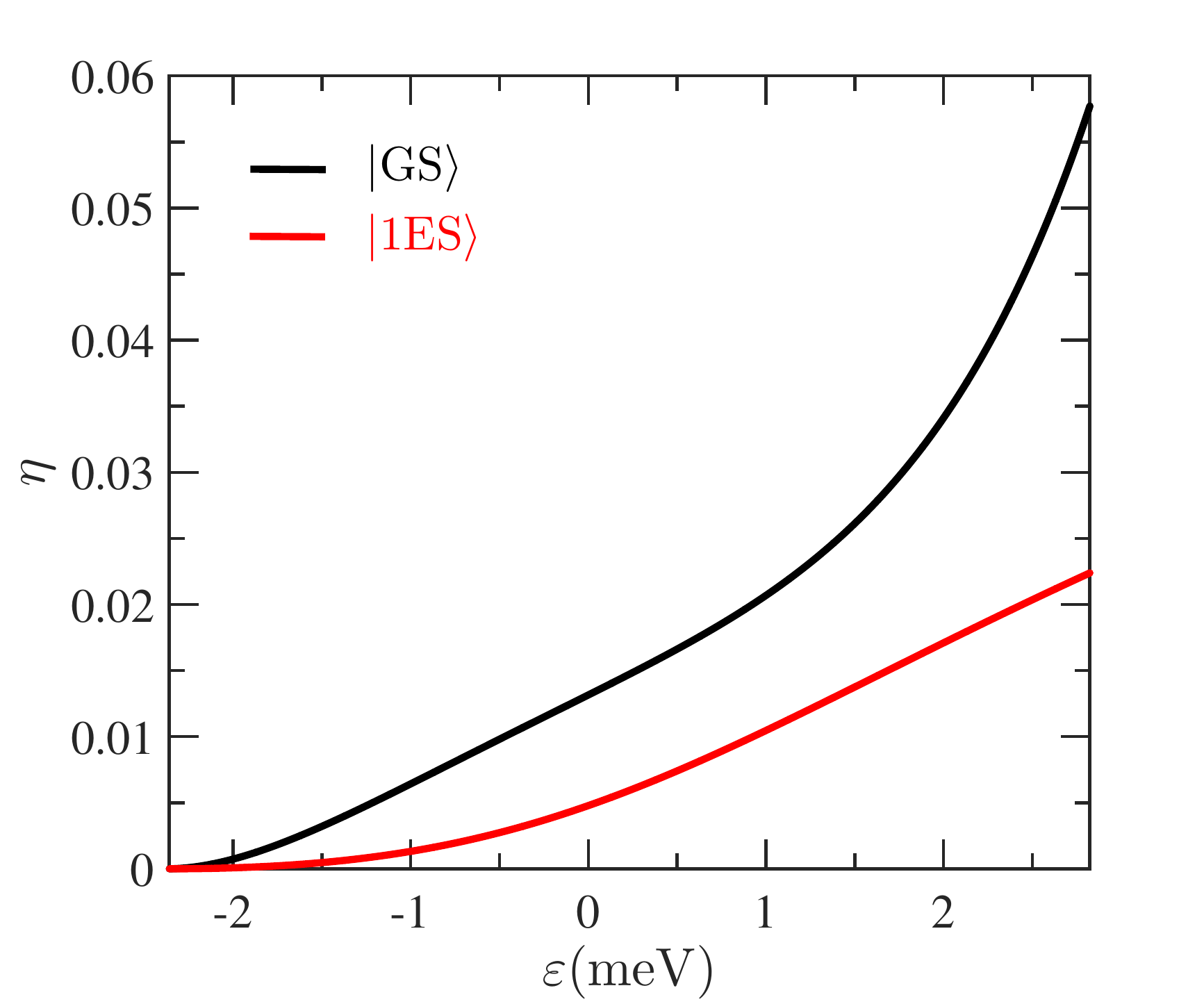}
\caption{The leakage parameter $\eta$ for both the ground state $\left|\mathrm{GS}\right\rangle$ and the first excited state $\left|\mathrm{1ES}\right\rangle$ v.s. the detuning $\varepsilon$ when the triple-dot system is operated as an EO qubit with $|0_L\rangle$ and $|1_L\rangle$ being qubit states. Parameters: $\hbar\omega_{0}=8 \mathrm{meV}$, $a=22 \mathrm{nm}$ and $\varepsilon^{m}=3.3 \mathrm{meV}$.}
\label{fig:8}
\end{figure}

Fig.~\ref{fig:8} shows the leakage parameter $\eta$, which is the probability of finding neither $|0_L\rangle$ nor $|1_L\rangle$ for the states concerned,  as  functions of detuning $\varepsilon$. The black line shows the results for $|\mathrm{GS}\rangle$ and the red(gray) line for $|\mathrm{1ES}\rangle$. $\eta=0$ for $\varepsilon=\varepsilon_L$, and increases as $\varepsilon$ is detuned. We note, however, that the leakage is significant when the exchange-only qubit is operated by a rotation around the 120$^\circ$ axis at $\varepsilon=\varepsilon_R$. For $|\mathrm{GS}\rangle$ $\eta$ is almost 6\% and for $|\mathrm{1ES}\rangle$ $\eta>2\%$. Traditionally it is believed that the difficulty of performing universal gates on an exchange-only qubit mainly arises from nuclear or charge noises. Our results, on the other hand, indicates that the leakage out of the computational space is also playing a key role in the decoherence.

\begin{figure}
\includegraphics[width=1.05\columnwidth]{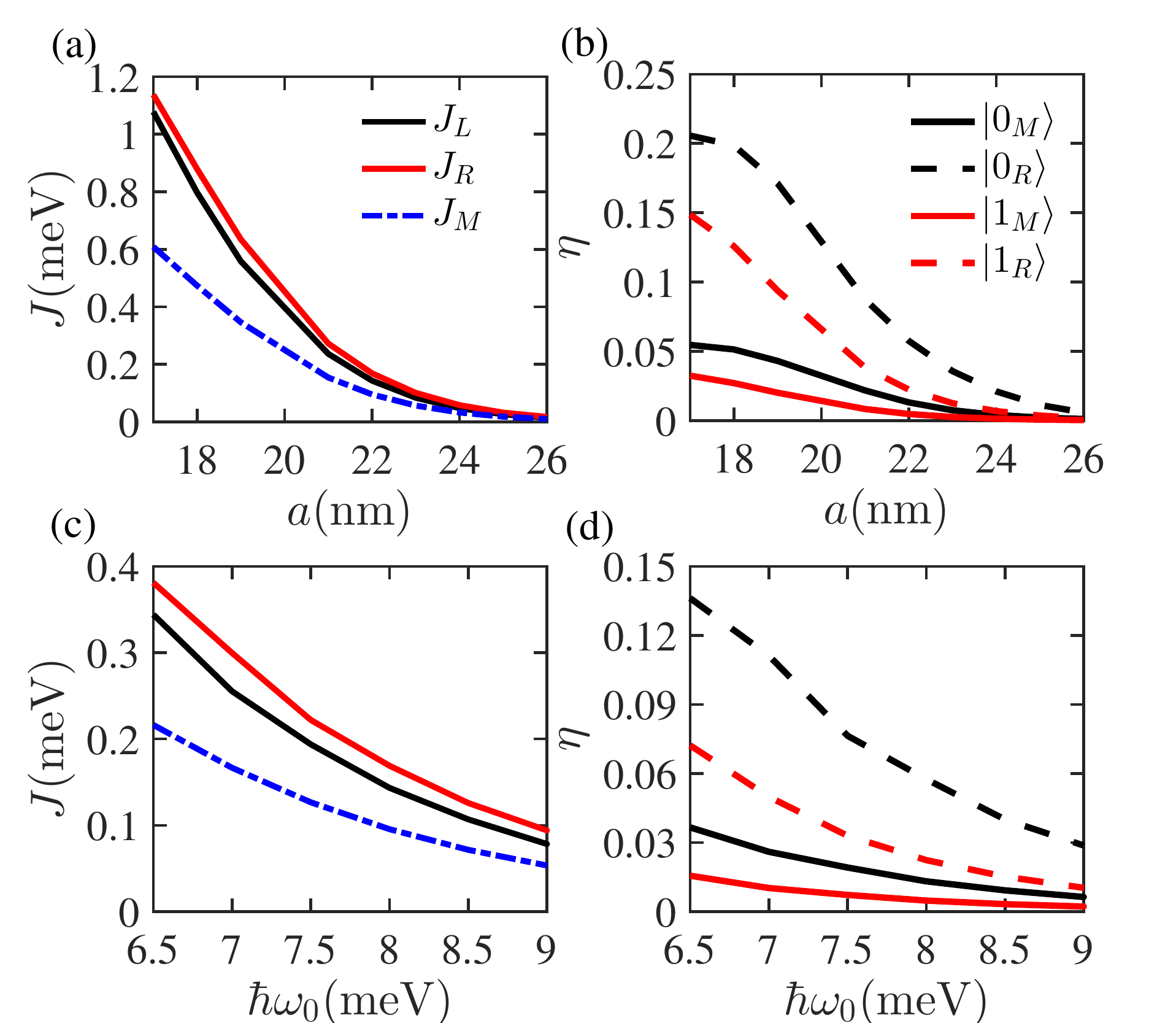}
\caption{ The exchange interaction and leakage as functions of the half inter-dot distance $a$ and the dot size $\hbar\omega_0$. For the exchange interactions, the results of $J_L$ (black solid lines), $J_R$ (red/gray solid lines) and $J_M$ (blue/gray dash-dotted lines) are shown. For the leakage, the results for $|0_M\rangle$ (black solid lines), $|0_R\rangle$ (black dashed lines), $|1_M\rangle$ (red/gray solid lines), and $|1_R\rangle$ (red/gray dashed lines) are shown. (Note that $\eta=0$ for $|0_L\rangle$  and $|1_L\rangle$. $\varepsilon^{m}$ is fixed on 3.3meV for all the cases.}
\label{fig:9}
\end{figure}

To study how one may suppress the leakage, we have calculated the exchange interactions and leakage as functions of the half inter-dot distance $a$ and the dot size $\hbar\omega_0$, and these results are presented in Fig.~\ref{fig:9}. The exchange interactions are calculated at three points: $J_L$ at $\varepsilon_L$, $J_M$ at $\varepsilon_M$, and $J_R$ at $\varepsilon_R$. From Fig.~\ref{fig:9}(a) and (c) we see that $J_L$ and $J_R$ are similar in amplitude and $J_R$ is typically a bit larger, which is due to the fact that $|\varepsilon_R|>|\varepsilon_L|$. $J_M$ is smaller than $J_L$ and $J_R$. In all cases, the exchange interaction decreases as either $a$ or $\hbar\omega_0$ is increased. Fig.~\ref{fig:9}(b) and (d) shows the results for the leakage. The leakage parameter, $\eta$ is much smaller for the states in the middle, $|0_M\rangle$ and $|1_M\rangle$, as compared to those on the right, 
$|0_R\rangle$ and $|1_R\rangle$. (Note that the leakage for $|0_L\rangle$ and $|1_L\rangle$ are zero because they are defined as the qubit states). We also see that while the leakage can be reduced by increasing $a$ or $\hbar\omega_0$, the exchange interactions decrease at the same time, making the gate operations slow. We therefore conclude that the problem of leakage during the operation of an exchange-only qubits around the 120$^\circ$ axis cannot be circumvented by enlarging either the interdot distance or the dot size, because at the same time the exchange interaction would be unacceptably slow.

\subsection{Resonant-exchange qubit}
\label{sec:rx}

A triple-quantum-dot system can alternatively be operated as an RX qubit, in which the $|0_M\rangle$ and $|1_M\rangle$ states are designated as the qubit states, and the qubit is operated in a small neighborhood of $\varepsilon=0$ by rf pulses \cite{Medford.13b}. The original motivation of the introduction of RX qubit is that 
$\varepsilon=0$ is a sweet spot of $J(\varepsilon)$ and operations based at this location is basically free of charge noise, while at the same time operations of an exchange-only qubit suffers from the charge noise. From the results of Sec.~\ref{sec:eo} we see that an additional disadvantage of the exchange-only qubit is that the leakage is significant when it is operated at the 120$^\circ$ axis. In this section, we discuss the sweet spot and leakage of an RX qubit and shall see that the leakage is negligible when the triple-quantum-dot system is operated as an RX qubit, which is another advantage of the RX operating scheme.

\begin{figure}
\includegraphics[width=0.9\columnwidth]{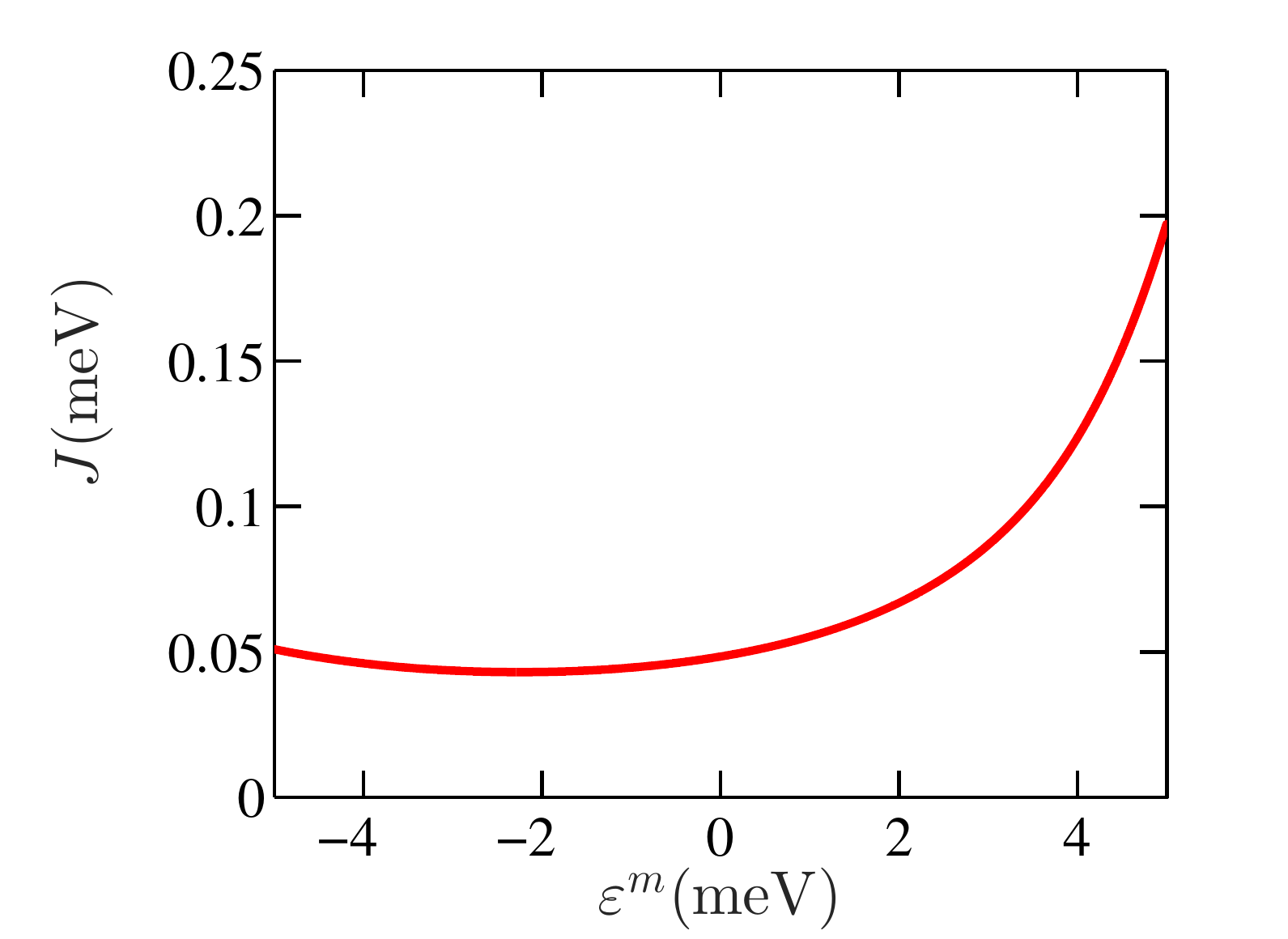}
\caption{The exchange interaction $J$ as a function of detuning $\varepsilon^{m}$. We can see that the sweet spot is at $\varepsilon^{m}=-2.25$ meV. Parameters: $\hbar\omega_{0}=8 \mathrm{meV}$, $a=22 \mathrm{nm}$ and $\varepsilon=0$.}
\label{fig:10}
\end{figure}

In Fig.~\ref{fig:10} we show the exchange interaction as a function of the detuning $\varepsilon^{m}$ while $\varepsilon$ is fixed at 0, namely $J(\varepsilon=0, \varepsilon^{m})$. We see that the sweet spot is at $\varepsilon^{m}=-2.25$ meV for this set of parameters. The exchange interaction increases more rapidly for $\varepsilon^{m}>-2.25$ meV than  $\varepsilon^{m}<-2.25$ meV.

\begin{figure}
\includegraphics[width=0.8\columnwidth]{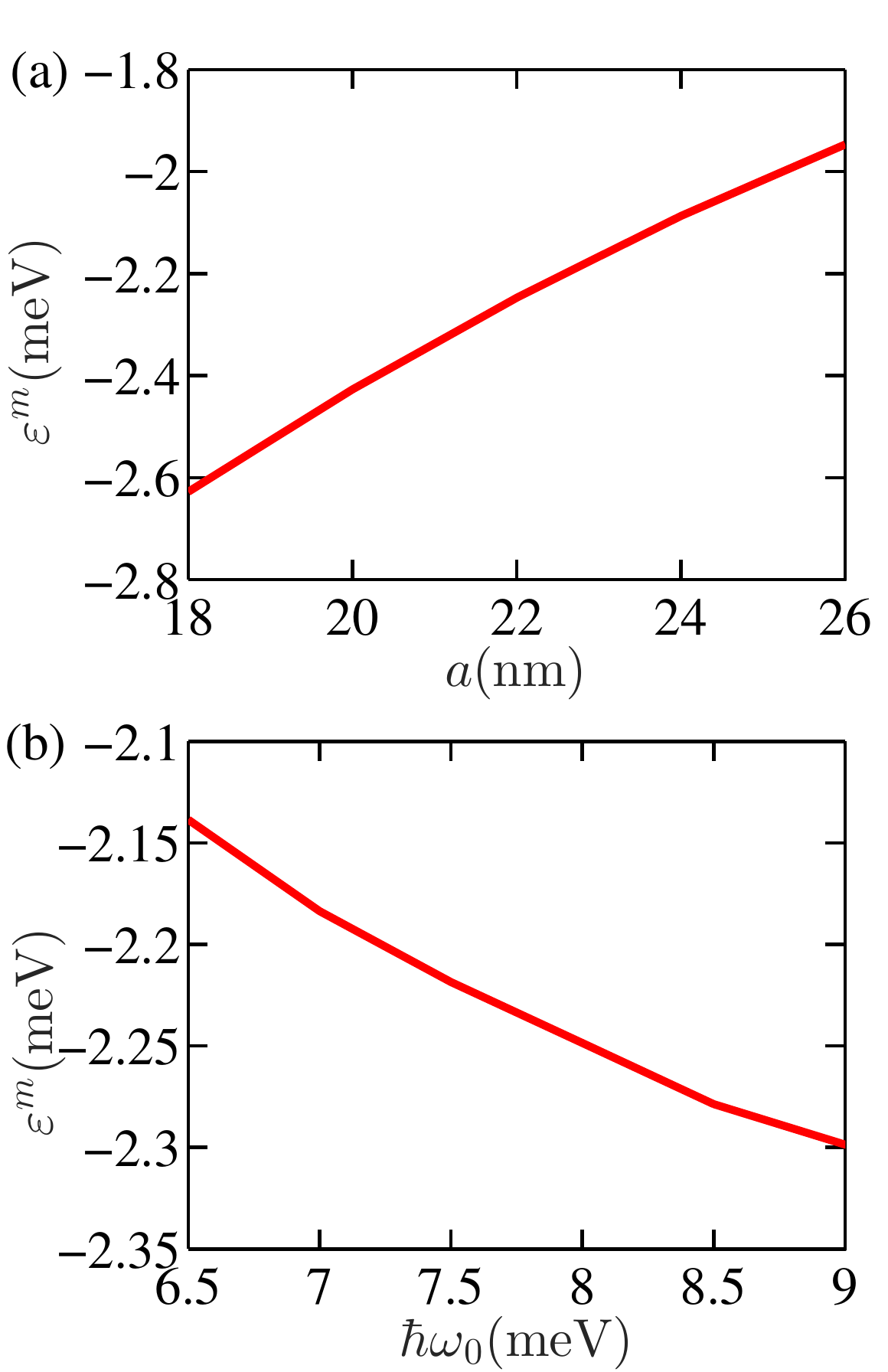}
\caption{The location of the sweet spot $\varepsilon^{m}$ as functions of the half-dot-distance $a$ [panel (a)] and the dot size $\hbar \omega_{0}$ [panel (b)]. Parameters: $\hbar\omega_{0}=8 \mathrm{meV}$, $a=22 \mathrm{nm}$ and $\varepsilon=0$.}
\label{fig:11}	
\end{figure}

In Fig.~\ref{fig:11} we study the locus of the sweet spot $\varepsilon^m$ as functions of $a$ and $\hbar\omega_0$. For relevant values of $18\mathrm{nm}\le a \le 26\mathrm{nm}$, $\varepsilon^m$ ranges between approximately -2.6 meV and -2 meV. The value of $\varepsilon^m$ increases as $a$ is increased such that the two dots are further apart from each other. Moreover,  $\varepsilon^m$ decreases as the dots get larger ($\hbar\omega_0$ is increased). These informations should be useful when one would like to fine tune the location of the sweet spot to maximize the performance of the qubit.

\begin{figure}
\includegraphics[width=0.75\columnwidth]{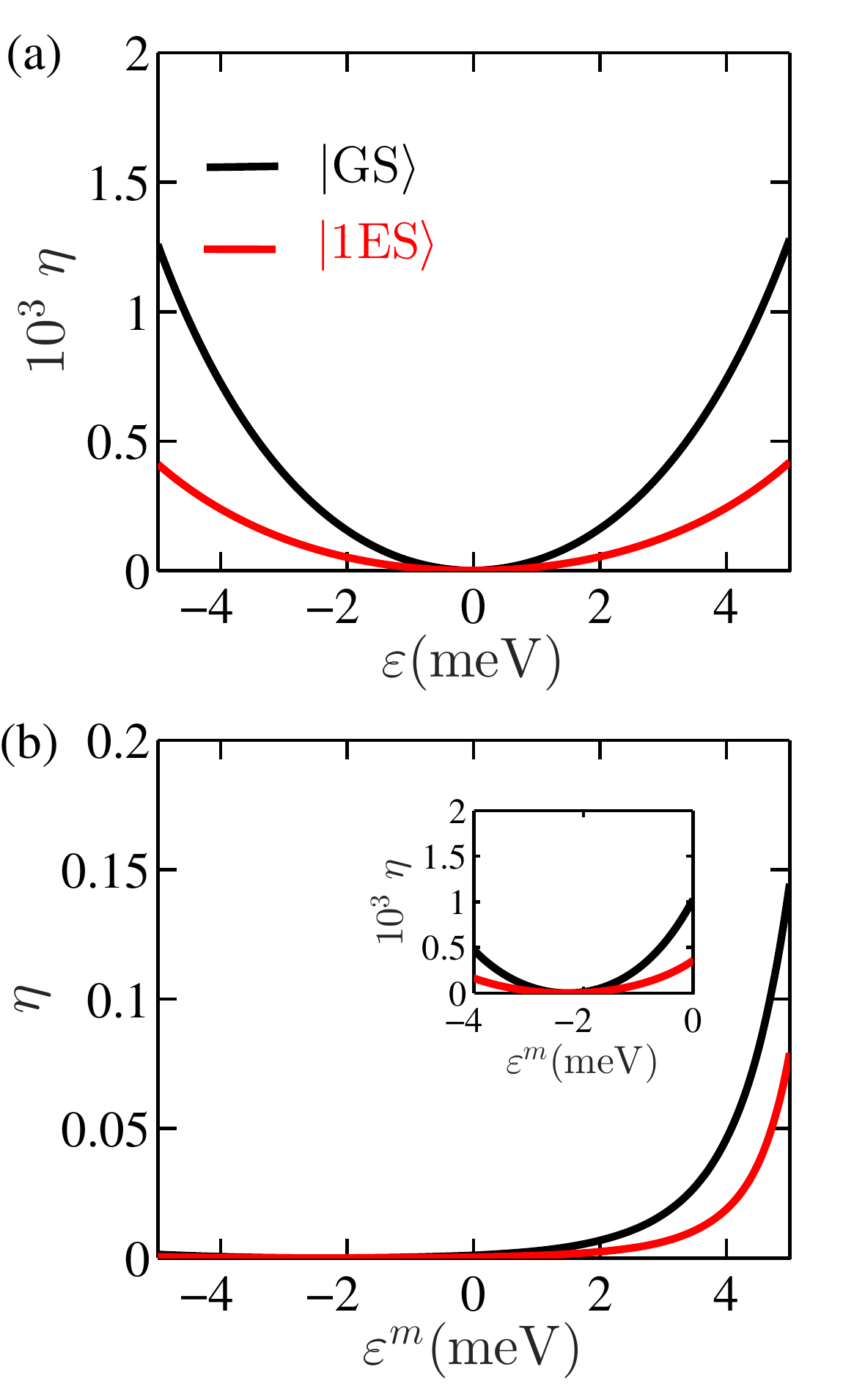}
\caption{The leakage parameter $\eta$ for a triple-dot system operated as an  RX qubit as functions of the detuning values. Panel (a): $\eta$ v.s. $\varepsilon$ while $\varepsilon^{m}$ is fixed at its sweet spot $\varepsilon^{m}=-2.25$  meV. Panel (b): $\eta$ v.s. $\varepsilon^{m}$ while $\varepsilon$ is fixed at its sweet spot $\varepsilon=0$. The black lines indicate the results for the $\left|\mathrm{GS}\right\rangle$ state while the red/gray lines the  $\left|\mathrm{1ES}\right\rangle$ state. Parameters: $\hbar\omega_{0}=8 \mathrm{meV}$, $a=22 \mathrm{nm}$.}
\label{fig:12}
\end{figure}

We now discuss the leakage of the triple-quantum-dot system when it is operated as an RX qubit. In Fig.~\ref{fig:12}(a) we show the leakage parameter $\eta$ as a function of the detuning $\varepsilon$ while the other detuning value $\varepsilon^{m}$ is fixed at its sweet spot $\varepsilon^{m}=-2.25$  meV. There is no leakage when $\varepsilon=0$, but application of the rf pulse will deviate from this point and we would like to study the leakage when $\varepsilon$ is away from the equilibrium point. From Fig.~\ref{fig:12}(a) we see that even if $\varepsilon$ is relatively far away from the  equilibrium point, the leakage remains small, which is of the order of $10^{-3}$, much smaller than the EO case. In Fig.~\ref{fig:12}(b), we fix $\varepsilon$ at its sweet spot $\varepsilon=0$, and vary $\varepsilon^m$. We see that the leakage is negligibly small in a reasonably wide neighborhood around the sweet spot $\varepsilon^m= -2.25$ meV, but increases rapidly when $\varepsilon^m\gtrsim 2$ meV. Since the rf control of an RX qubit is always within a small neighborhood of the equilibrium point, our results indicate that the leakage remains small in this case as long as the equilibrium point is chosen at or close to the double-sweet-spot point. The fact that leakage can be substantially suppressed is another advantage of RX qubit as compared to the EO one. This reduction of leakage can be understood as follows. Firstly, in the operation of an RX qubit, the detuning need not be varied much, so the deviation from the qubit states are minimal, much smaller than the case of the EO qubit in which one has to tune $\varepsilon$ over a range of several meV. Second, from Fig.~\ref{fig:rx} we see that $t_{12}$ is very small for $\varepsilon<0$, so leakage to the $(2,0,1)$ state is greatly suppressed. Similarly, $t_{23}$ is small for $\varepsilon>0$, which reduces leakage to the $(1,0,2)$ state.  Nevertheless, for an EO qubit it is necessary to be able to detune to $(2,0,1)$ and $(1,0,2)$ states for initialization, readout and also for operations of the qubit, so that a positive $\varepsilon^m$ is favored. Unfortunately, in this case the leakage to the $(2,0,1)$ and $(1,0,2)$ states is  considerable as demonstrated in Fig.~\ref{fig:eo}. When the system is detuned positively, $\epsilon_3$ that is already smaller than $\epsilon_2$ is further reduced, and an electron in dot 2 is more likely to join dot 3 so that leakage to the $(1,0,2)$ state is enhanced. Therefore the RX qubit gains this advantage of having  reduced leakage because it allows more freedom in choosing the detuning values, especially $\varepsilon^m$, and also due to the fact that varying the detuning over a small range near the sweet spot suffices for qubit operation.

\begin{figure}
\includegraphics[width=1.05\columnwidth]{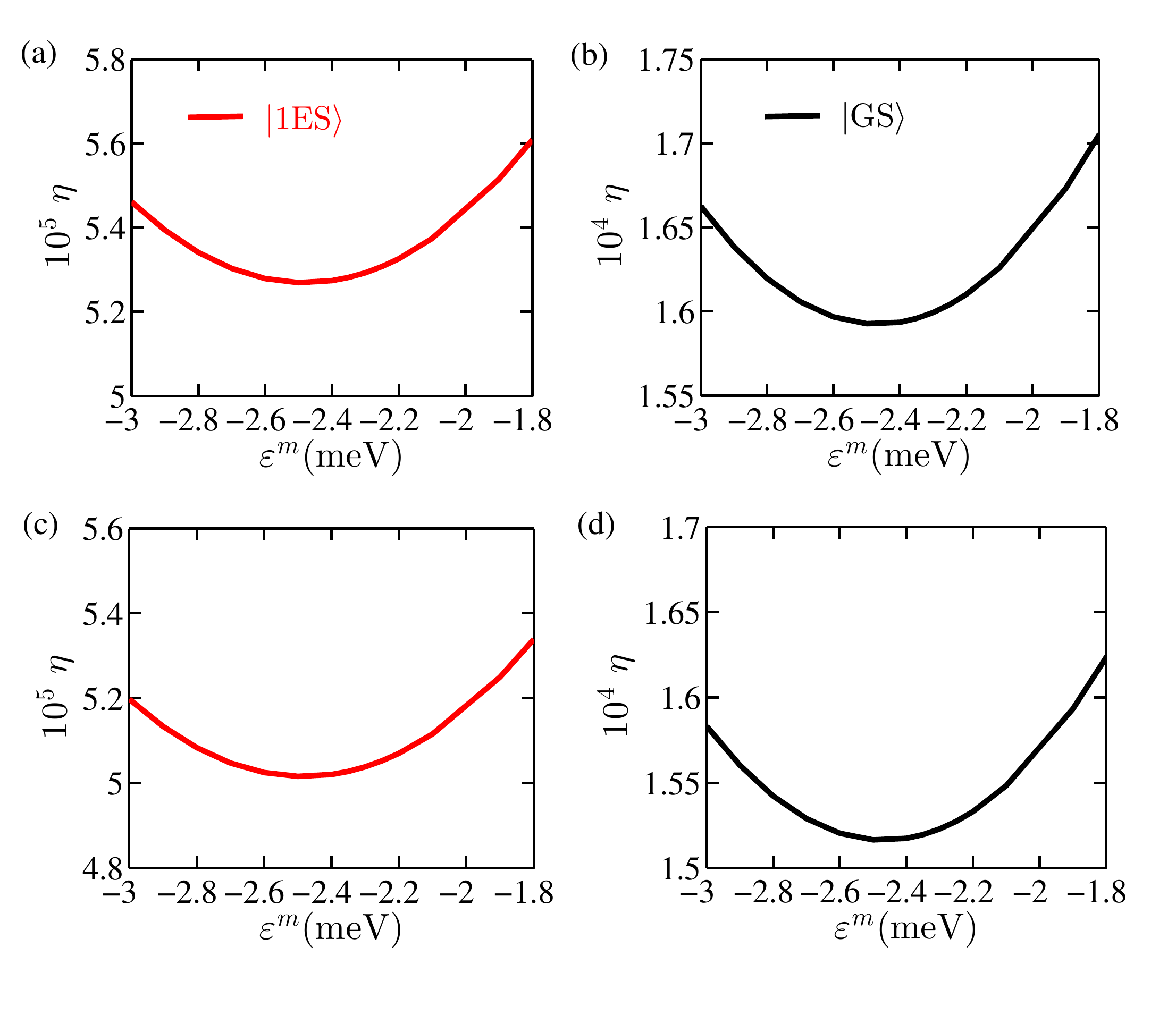}
\caption{
The leakage parameter $\eta$ for a triple-dot system operated as an  RX qubit as functions of the detuning $\varepsilon^{m}$. For (a) and (b), $\varepsilon= 2 \mathrm{meV}$; for (c) and (d), $\varepsilon= -2 \mathrm{meV}$ (away from the sweet spot $\varepsilon=0$) The black lines indicate the results for the $\left|\mathrm{GS}\right\rangle$ state while the red/gray lines the  $\left|\mathrm{1ES}\right\rangle$ state. Parameters: $\hbar\omega_{0}=8 \mathrm{meV}$, $a=22 \mathrm{nm}$.
} 
\label{fig:13}	
\end{figure}

\begin{figure}
\includegraphics[width=1.05\columnwidth]{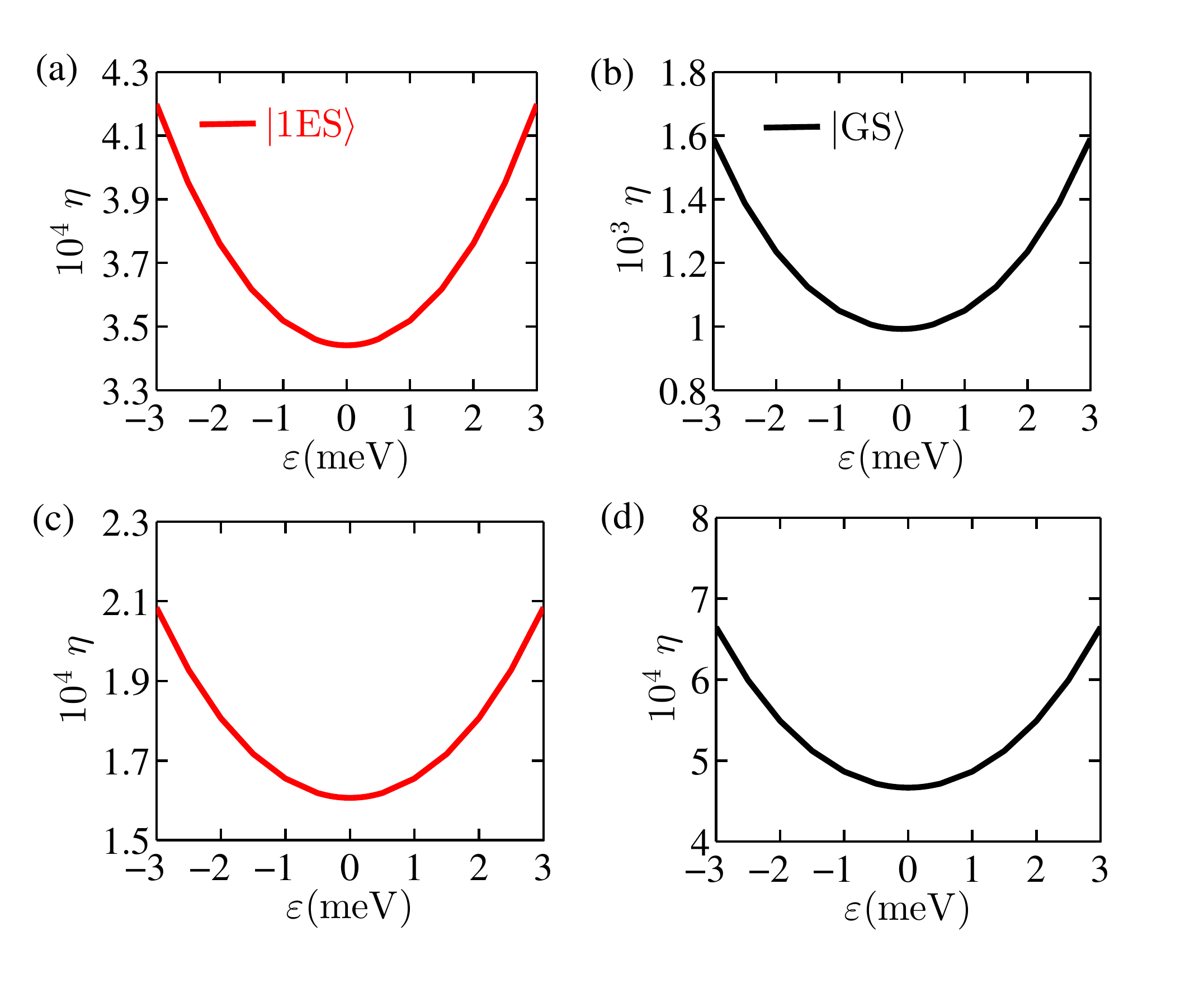}
\caption{
The leakage parameter $\eta$ for a triple-dot system operated as an RX qubit as functions of the detuning $\varepsilon$. For (a) and (b), $\varepsilon^{m}=0 $; for (c) and (d), $\varepsilon^{m}= -4 \mathrm{meV}$ (away from the sweet spot $\varepsilon^m=-2.25$ meV) The black lines indicate the results for the $\left|\mathrm{GS}\right\rangle$ state while the red/gray lines the  $\left|\mathrm{1ES}\right\rangle$ state. Parameters: $\hbar\omega_{0}=8 \mathrm{meV}$, $a=22 \mathrm{nm}$.
}
\label{fig:14}	
\end{figure}

In Figs.~\ref{fig:13} and \ref{fig:14} we provide further results on whether the double sweet spot, where the charge noise is at minimum, is also  close to the point where the leakage is minimum. In Fig.~\ref{fig:13} we study how the leakage varies with the detuning $\varepsilon^m$ and have chosen two values of $\varepsilon$ symmetrically from both sides of the sweet spot $\varepsilon=0$, i.e. $\varepsilon=\pm2$ meV. This is to simulate an rf operation around $\varepsilon=0$ with amplitude 2 meV, and at $\varepsilon=\pm2$ meV the leakage should take its maximal value in the course of such operation. We can see from Fig.~\ref{fig:13} that the leakage is minimal at $\varepsilon^m\approx-2.45$ meV. Although it is not exactly at the sweet spot $-2.25$ meV, it is rather close and the leakage would not increase much at the sweet spot. In Fig.~\ref{fig:14} we fix $\varepsilon^m$ to two values on both sides of the sweet spot, 0 and $-4$ meV. Not surprisingly, the leakage is minimal at the symmetric point, $\varepsilon=0$.   We also see that the leakage for the ground state is typically larger than that of the first excited state by a factor of 3 to 5.

\section{Conclusions}
\label{sec:conclusion}
In this paper, we have performed a molecular-orbital calculation of a triple-quantum-dot system under the Hund-Mulliken approximation. We have taken a total of nine three-electron states into consideration, including those involving double occupancy in one of the quantum dots. We have taken the external magnetic field as zero so that there is no leakage to the $|Q\rangle$ state. Nevertheless, leakage to other states with double occupancy does happen. Our calculation indicates that when the triple-quantum-dot system is treated as an EO qubit, leakage is significant when rotating around the axis 120$^\circ$ from $\hat{z}$. While this leakage can be suppressed by either enlarging the inter-dot distance or shrinking each quantum dot, the exchange interaction is also reduced at the same time, making the gate operations unacceptably slow. Alternatively, when the same triple-quantum-dot system is operated as an RX qubit, the leakage can be almost completely suppressed since the rf pulse only operates in a small neighborhood around the equilibrium point, where leakage is zero by definition. Even considering the maximal leakage while the rf pulse brings the detuning away form the equilibrium point, the leakage is at least two order of magnitudes smaller than that of the EO qubit case. We have also calculated the location of the sweet-spot $\varepsilon^m$ as functions of the inter-dot distance and the dot size, and have found that $\varepsilon^m$ becomes more negative either when the dots are closer, or when each of the dots is smaller in size. In addition, we calculated how the leakage depends on the values of the detuning. While $\varepsilon^m$ deviates from the equilibrium value, the leakage is smallest at the symmetric point $\varepsilon=0$, which is at the same time the sweet spot. On the other hand, for the detuning $\varepsilon^m$, although the point where leakage is smallest is not exactly the sweet-spot point, they are rather close. Our results indicate that it should be optimal to operate the RX qubit at the double-sweet-spot point, both for the purpose of reducing charge noise and suppressing leakage to other states.

This work is supported by the 
Research Grants Council of the Hong Kong Special Administrative Region, China (No.~CityU 21300116, CityU 11303617), the National Natural Science Foundation of China (No.~11604277), and the Guangdong Innovative and Entrepreneurial Research Team Program (No.~2016ZT06D348)


%

\end{document}